\documentclass[11pt,a4paper]{article}

\usepackage[T1]{fontenc}
\usepackage[latin1]{inputenc}
\usepackage{amsmath,amssymb,amsthm,amscd}
\usepackage{graphics}
\usepackage{epic}
\usepackage{epsfig}
\usepackage{bbm}
\usepackage{tikz}
\usetikzlibrary{shapes}
\usepackage{stmaryrd}
\usepackage{hyperref}
\usepackage{subcaption}
\usepackage{natbib}
\numberwithin{equation}{section}

\def\11{\mathbbm{1}}

\newtheorem{thm}{Theorem}[section]

\newtheorem{prop}[thm]{Proposition}

\begin{document}

\title{Modelling the effects of biological intervention in a dynamical gene network}

\author{Nicolas Champagnat$^{1,*}$, Rodolphe Loubaton$^{2,3}$,\\ Laurent Vallat$^{4,5}$, Pierre Vallois$^{1}$}

\footnotetext[1]{Universit\'e de Lorraine, CNRS, Inria, IECL, F-54000 Nancy, France\\ * Corresponding author: Nicolas.Champagnat@inria.fr}
\footnotetext[2]{VetAgro Sup, d\'epartement  Territoires et Soci\'et\'e, CS 82212, F-63370 Lempdes, France}
\footnotetext[3]{INRAe, \'equipe CARAIBE, UMR 1213, F-63122 St Genes Champanelle, France}
\footnotetext[4]{University of Strasbourg, CNRS, UMR-7242 Biotechnology and Cell Signaling, F-67400 Illkirch, France}
\footnotetext[5]{Department of Molecular Genetic of Cancers, Strasbourg University Hospital, F-67200 Strasbourg, France}

% \date{}

\maketitle

\begin{abstract}
  Cellular response to environmental and internal signals can be modeled by dynamical gene regulatory networks (GRN). In the
  literature, three main classes of gene network models can be distinguished: (i) non-quantitative (or data-based) models which do
  not describe the probability distribution of gene expressions; (ii) quantitative models which fully describe the probability
  distribution of all genes co-expression; and (iii) mechanistic models which allow for a causal interpretation of gene interactions.
  We propose two rigorous frameworks to model gene alteration in a dynamical GRN, depending on whether the network model is
  quantitative or mechanistic. We explain how these models can be used for design of experiment, or, if additional alteration data
  are available, for validation purposes or to improve the parameter estimation of the original model. We apply these methods
  to % several examples of GRN, including
  the Gaussian graphical model, which is quantitative but non-mechanistic, and to mechanistic models of Bayesian networks and
  penalized linear regression.
\end{abstract}

% \noindent\emph{MSC 2000 subject classification:} Primary: ; secondary: .


\medskip


\section{Introduction}

The functioning of eukaryotic cells in their basal state, and their response to changes in their environment, are dynamic phenomena
involving multiple players activated in a cascade with numerous levels of regulation. The sequential activation of multiple genes
(transcriptional response) following an extra-cellular signal is a crucial step in this response. Some of these genes are
transcriptional factors which activate other genes in cascade, notably genes encoding proteins which are the effectors of the
cellular response adapted to the initial signal.

These various interactions between genes can be represented by gene interaction networks (GRNs)~\citep{barabasi2004, Filkov2005,
  Hecker2009, Natale2018, Huynh2019survey}.

Recent developments in high-throughput sequencing techniques (RNA-seq) have made it possible to identify and quantify the various
players (messenger RNAs) in these interaction networks, at different times after cell activation~\citep{Schleiss2021}. These
sequencing data (dataset) can then be used to infer a mathematical model of the underlying dynamic gene interaction network.

\begin{figure}
  % \captionsetup[subfigure]{justification=centering}
  % \begin{subfigure}{0.35\textwidth}
  \begin{center}
    \begin{tikzpicture}
      \draw (-4,0.5) node {(a)};
      \node[draw,rectangle,rounded corners=3pt] (G1) at (-3,0) {$\mu_1,\sigma_{11}$};
      \node[draw,rectangle,rounded corners=3pt] (G2) at (0,0) {$\mu_2,\sigma_{22}$};
      \node[draw,rectangle,rounded corners=3pt] (G3) at (-3,-2) {$\mu_3,\sigma_{33}$};
      \node[draw,rectangle,rounded corners=3pt] (G4) at (0,-2) {$\mu_4,\sigma_{44}$};
      \draw (G1) -- (G2);
      \draw (-1.5,0) node [above] {$\sigma_{12}$};
      \draw (G2) -- (G4);
      \draw (0,-1) node [right] {$\sigma_{24}$};
      \draw (G1) -- (G3);
      \draw (-3,-1) node [left] {$\sigma_{13}$};
      \draw (G3) -- (G4);
      \draw (-1.5,-2) node [below] {$\sigma_{34}$};
      \draw (G1) -- (G4);
      \draw (-2.35,-0.5) node [below] {$\sigma_{14}$};
      \draw (G3) -- (G2);
      \draw (-0.6,-0.5) node [below] {$\sigma_{23}$};
      \draw (2,0.5) node {(b)};
      \node[draw,rectangle,rounded corners=3pt] (G1) at (3,0) {$0,0$};
      \node[draw,rectangle,rounded corners=3pt] (G2) at (6,0) {$\mu'_2,\sigma'_{22}$};
      \node[draw,rectangle,rounded corners=3pt] (G3) at (3,-2) {$\mu'_3,\sigma'_{33}$};
      \node[draw,rectangle,rounded corners=3pt] (G4) at (6,-2) {$\mu'_4,\sigma'_{44}$};
      \draw (G2) -- (G4);
      \draw (6,-1) node [right] {$\sigma'_{24}$};
      \draw (G3) -- (G4);
      \draw (4.5,-2) node [below] {$\sigma'_{34}$};
      \draw (G3) -- (G2);
      \draw (5.4,-0.5) node [below] {$\sigma'_{34}$};
    \end{tikzpicture}
  \end{center}
  \caption{Schematic representation of a Gaussian graphical model and prediction of alteration (knock-out) of gene 1. Each node
    corresponds to a gene labeled from 1 to 4. (a) Without alteration: the mean expression of gene $i$ is $\mu_i$ and the covariance
    matrix of genes expression is $(\sigma_{ij})_{1\leq i,j\leq 4}$. (b) After knock-out of gene 1: $\mu_1$ and $\sigma_{11}$ were
    set to 0; the question of prediction amounts to determine the new values of the means and covariances $(\mu'_i)_{2\leq i\leq 4}$
    and $(\sigma'_{ij})_{2\leq i,j\leq 4}$.}
  \label{fig:GGM}
\end{figure}
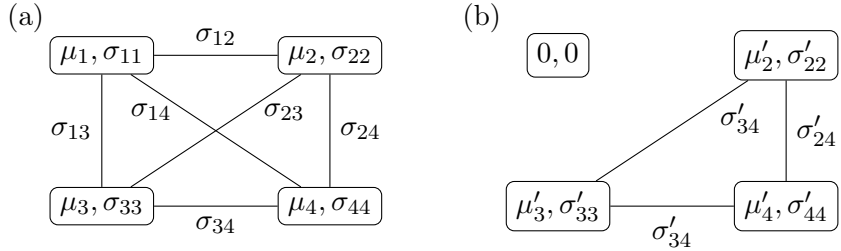

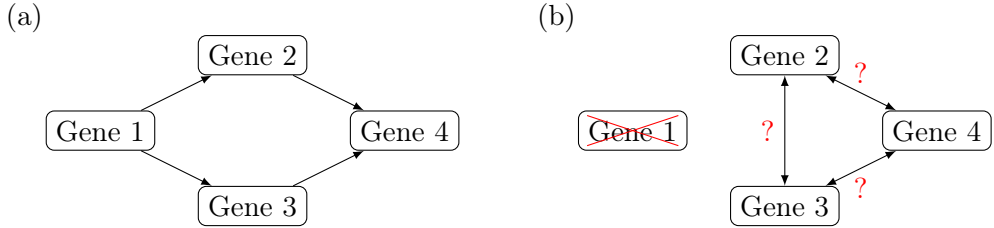
\begin{figure}
  % \captionsetup[subfigure]{justification=centering}
  % \begin{subfigure}{0.35\textwidth}
  \begin{center}
    \begin{tikzpicture}
      \draw (-4,0.5) node {(a)};
      \node[draw,rectangle,rounded corners=3pt] (G1) at (-3,-1) {Gene 1};
      \node[draw,rectangle,rounded corners=3pt] (G2) at (-1,0) {Gene 2};
      \node[draw,rectangle,rounded corners=3pt] (G3) at (-1,-2) {Gene 3};
      \node[draw,rectangle,rounded corners=3pt] (G4) at (1,-1) {Gene 4};
      \draw[->,>=latex] (G1) -- (G2);
      % \draw (-1.5,0) node [above] {$\sigma_{12}$};
      \draw[->,>=latex] (G1) -- (G3);
      % \draw (0,-1) node [right] {$\sigma_{24}$};
      \draw[->,>=latex] (G2) -- (G4);
      % \draw (-3,-1) node [left] {$\sigma_{13}$};
      \draw[->,>=latex] (G3) -- (G4);
      % \draw (-1.5,-2) node [below] {$\sigma_{34}$};
      % \draw[<->,>=latex] (G1) -- (G4);
      % \draw (-2.35,-0.5) node [below] {$\sigma_{14}$};
      % \draw[<->,>=latex] (G3) -- (G2);
      % \draw (-0.6,-0.5) node [below] {$\sigma_{34}$};
      \draw (3,0.5) node {(b)};
      \node[draw,rectangle,rounded corners=3pt] (G1) at (4,-1) {Gene 1};
      \draw[color=red] (3.4,-1.2) -- (4.6,-0.8);
      \draw[color=red] (3.4,-0.8) -- (4.6,-1.2);
      \node[draw,rectangle,rounded corners=3pt] (G2) at (6,0) {Gene 2};
      \node[draw,rectangle,rounded corners=3pt] (G3) at (6,-2) {Gene 3};
      \node[draw,rectangle,rounded corners=3pt] (G4) at (8,-1) {Gene 4};
      % \draw[->,>=latex] (G1) -- (G2);
      % \draw (-1.5,0) node [above] {$\sigma_{12}$};
      % \draw[->,>=latex] (G1) -- (G3);
      % \draw (0,-1) node [right] {$\sigma_{24}$};
      \draw[<->,>=latex] (G2) -- (G4);
      \draw[color=red] (7,-0.5) node [above] {?};
      \draw[<->,>=latex] (G3) -- (G4);
      \draw[color=red] (7,-1.5) node [below] {?};
      \draw[<->,>=latex] (G2) -- (G3);
      \draw[color=red] (6,-1) node [left] {?};
    \end{tikzpicture}
  \end{center}
  \caption{Schematic representation of a mechanistic model and prediction of alteration (knock-out) of gene 1. Each node corresponds
    to a gene labeled from 1 to 4. (a) Without alteration: arrows represent direct influence of a gene on another; they encode a
    quantitative relation (not shown here). (b) After knock-out of gene 1: the problem of prediction consists in determining how the
    arrows and the quantitative relations they encode are modified.}
  \label{fig:meca}
\end{figure}

Gene regulatory network reconstruction from gene expression data is a difficult and ill-posed problem~\citep{Minhaz2014}. A large
variety of methods were proposed, which may give different interpretations to the edges of the infered
network~\citep{Hecker2009,Wang2014,Natale2018,Huynh2019survey}. %,Grzegorczyk2019}.
These methods can be classified into three main categories: (i) \emph{data-driven methods} which do not model gene expression and
only represent interactions between genes from statistical quantities such as correlation or mutual information; (ii)
\emph{quantitative methods} which specify a parametric family of distributions for the genes expressions; (iii) \emph{mechanistic
  methods} which are particular cases of quantitative methods where the model quantitatively represents the effects of interactions
between small subsets of genes (typically pairs of genes). Mechanistic models may have different levels of description of biological
interactions ranging from phenomenological, often indirect interactions to chemical direct interactions based on the biological
understanding of gene regulation (causal networks~\citep{pearl2016}). In this work, we focus on quantitative (mechanistic or
non-mechanistic) methods.
% In non-mechanistic models, the gene network may be directed or non-directed. For mechanistic models, it is typically directed.
The Gaussian graphical model and its variants are typical examples of quantitative but non-mechanistic models~\citep{Chiquet2019}. In
the Gaussian graphical model, the edges in the gene network are non-directed and the model is characterized by the vector of mean
expression $(\mu_i)_{1\leq i\leq N}$ of $N$ genes and their covariance matrix $(\sigma_{ij})_{1\leq i,j\leq N}$, as represented in
Fig.~\ref{fig:GGM}(a). Mechanistic models can range from simple models such as models based on penalized linear
regressions~\citep{Vallat2013,Omranian2016,Ishikawa2023} to complex models, e.g.\ based on ordinary differential
equations~\citep{Goutsias2007} or piecewise-deterministic models~\citep{Bonnaffoux2019,Ventre2023}. Other standard classes of
mechanistic models used for gene network reconstruction are Boolean networks~\citep{Ideker1999}, and Bayesian
networks~\citep{Chang2011}. In mechanistic models, the gene network is usually represented using an oriented graph as in
Fig.~\ref{fig:meca}(a). Another popular class of gene network models is based on random forests~\citep{Huynh2018}, but they are
usually difficult to interpret as quantitative model because they only give a phenomenological representation of interactions between
genes, so we will not consider them in this work.
% , although some studies address similar questions as ours for such models~\citep{??}

Certain situations can alter the structure or expression of a gene. For example, in pathological situation, an alteration (mutation,
deletion...) can occur in a gene and induce its loss of expression, leading to cell dysfunction and the development of an
inflammatory disease or cancer. In experimental situations, it is possible to alter (suppress, decrease or increase) the expression
of a specific gene involved in an interaction network, in order to better characterize its role in cell function. This experimental
alteration can target the structure of a particular gene to modulate (decrease or increase) its expression, or directly target the
expression of this gene inhibiting the messenger RNA (mRNA). These approaches are leading to the prospect of therapeutic applications
in humans~\citep{de2024}.

The CRISPR/CAS9 approach is now one of the most widely used methods for altering gene structure. This method is derived from a
natural antiviral defense mechanism found in certain bacteria. After a viral (phage) infection, these bacteria retain fragments of
this virus's sequences (CRISPR sequences) in their genome. Upon re-infection with this virus, the viral DNA is recognized by a
bacterial CRISPR RNA (crRNA). This crRNA then associates with a transactivating CRISPR RNA (tacrRNA) and recruits the CAS9
endonuclease, an enzyme that cleaves the viral sequence, inducing its degradation and the elimination of the
virus~\citep{boettcher2015}. Based on this mechanism, it is possible to synthesize a target gene-specific gRNA and associate it with
the CAS9 endonuclease. After introduction of this molecule into a eukaryotic cell, the gRNA directs the endonuclease towards the
target gene to induce its cleavage. During DNA repair mechanisms, a mutation altering the sequence of this gene is introduced, making
it possible to specifically inhibit expression of this gene~\citep{doudna2014,vercauteren2024}.

Some approaches directly target the expression (mRNA) of a particular gene in the GRN. These methods are also derived from
physiological mechanisms of gene expression regulation and natural cell defense mechanisms based on RNA interference by microRNA
(miRNA). When viral RNA is present in a cell, cellular miRNAs specifically recognize the sequence of these mRNAs. The miRNAs then
bind to these viral RNAs, recruit the RISC/AGO2 molecular complex and induce their degradation~\citep{lee1993}. The siRNA gene
silencing method uses the same principle. After design of a 20-24 nucleotide synthetic siRNA specific to the gene sequence to be
silenced, this siRNA is experimentally introduced into the cell. This siRNA then binds to the mRNA transcript of the target gene,
recruits the RISC complex and uses the cellular machinery to induce degradation of the target RNA
molecules~\citep{fire1998,davidson2011}. Based on the same principle, the shRNA silencing method can also be used. In this case, a
plasmid encoding a pre-shRNA is introduced into the cell and the latter matures into siRNA in the cell cytoplasm. This siRNA then
recruits the RISC complex to degrade its specific target RNA~\citep{mohr2014}.

Other experimental methods can be used to increase the expression of a particular gene in a cell. For example, techniques for
introducing recombinant DNA (the sequence of a gene) into a cell using a vector in order to induce expression of that gene and use
the cellular machinery to produce the proteins of interest in that cell have been used for many years in the
laboratory~\citep{wolff1990} and are applicable for eukaryotic cells~\citep{martinez2022}.

% {\color{magenta} [J'ai commenc\'e \`a traduire ici, car sont exprim\'es les trois objectifs, sans d\'eveloppement biologique]}

% [ {\color{blue} les commentaires sont en bleu. J'ai essay\'e de simplifier, \`a v\'erifier !}]

It is therefore essential to predict the % effects of this alteration on the gene network and on the
cellular response to a gene alteration in a network, in order to gain a better understanding of the development of pathologies caused
by these alterations and acquired during life (cancer) or inherited (congenital diseases)~\citep{Chang2011,Vallat2013,Qin2014}. A
model of gene alteration can also be used for design of experiment, for example to determine target genes to act on to obtain a
certain biological effect or to discriminate between several possible gene networks~\citep{Ideker1999,Tegner2003,Dehghannasiri2015,
  Vundavilli2019,Bouvier2024}. When data of a genetic alteration experiment are available, a model of this alteration can also
be used to validate or improve the statistical estimation of the the initial model and/or the alteration
model~\citep{Villa2014,Omranian2016,secilmis2022,sarmah2022,Ishikawa2023}.

Modeling studies of gene alteration have used a wide variety of methods and models that all belong to the framework of mechanistic
models (Boolean networks~\citep{Azuaje2010,Ideker1999}, Bayesian networks~\citep{Chang2011}, linear regression
models~\citep{Vallat2013,Qin2014,Villa2014,Omranian2016}, ODE-based models ~\citep{Tegner2003,Goutsias2007}). Some models are
static~\citep{Tegner2003,Qin2014}, or focus on the stationary state of the network~\citep{Goutsias2007}, or assume that data at different
times are independent~\citep{Omranian2016}. Other studies only take into account alteration of the initial state in a dynamical
model~\citep{Chang2011,Qin2014,Omranian2016}. The most advanced models are those that consider dynamic models and take into account
changes in model structure over time~\citep{Vallat2013,Villa2014,Dehghannasiri2015, Vundavilli2019,Ishikawa2023,Bouvier2024}.

% In such a diverse context, it seems necessary to provide a unified framework for all these approaches.
Our aim is to define general methods to construct dynamical models of gene alteration that are applicable to both quantitative and
mechanistic settings and that are not limited to the simple change of initial condition.
% . The methods we propose allow us to account for how an alteration can go beyond the simple modification of the
% initial condition by modifying the model over time.
We set ourselves three goals: \emph{i)} understanding how gene alteration modifies cellular behavior, \emph{ii)} experiment design
and \emph{iii)} answering statistical issues such as estimation or validation.

We propose two different approaches, depending on whether the model is quantitative or mechanistic. In a so-called \emph{conditional
  approach}, we define an alteration as a probabilistic continioning of the stochastic initial model by a certain event, so that for
example the total knock-out of a gene is taken into account by zeroing its expression. In particular, we show that this genetic
alteration transforms a Gaussian graphical model into another Gaussian model whose mean values and correlations between gene
expressions can be explicitly calculated, see Fig.~\ref{fig:GGM}(b). In a \emph{mechanistic approach}, alteration takes the form of a
modification of the model's parameters, so that for example the total knock-out of a gene modifies the gene network as illustrated in
Fig.~\ref{fig:meca}(b).

Conditional and mechanistic methods are described in Section~\ref{sec:construction-IB}. They predict the effect of a gene alteration
from an initial model inferred from a dataset obtained without gene alteration. In Section \ref{sec:choix-xp}, we develop a first
application of our modeling to the design of the most relevant alteration experiments (according to a certain criterion). When a gene
alteration dataset is available, it is used to estimate the parameters of the model describing the alteration and/or to improve the
estimation of the parameters of the initial model (Section~\ref{sec:stat}). We then describe these methods in detail within the
framework of three standard classes of gene network models: an example of quantitative but non-mechanistic model, the Gaussian
graphical models, in Section~\ref{sec:GGM}, and two examples of mechanistic models in Section~\ref{sec:GNI-mecaniste}: Bayesian
networks in Section~\ref{sec:bayes}, and a penalized linear regression model from~\citep{Rodolphe2025} in Section~\ref{sec:lasso}.

% Our approach can also be applied to the modeling of gene alteration within the
% framework of Boolean networks

% {\color{blue} Si on dit  qu'on ne le fait pas parce qu'il y a d\'ej\`a beaucoup d`\'etudes de mod\'elisation d'alt\'eration g\'eniques, cela va \`a l'encontre de  notre objectif d'"universalit\'e". Il faudrait un autre argument pour justifier le fait que l'on ne les d\'eveloppent pas ...}

%  sur
% ces mod\`eles~\citep{Ideker1999,Azuaje2010,Zhu2016}. 

\section{Modeling gene network alteration}
\label{sec:construction-IB}

% \textcolor{red}{\`A citer ?
%   \begin{itemize}
%   \item   Effects of a silenced gene in Boolean network models
% Emir Haliki, Nadide Kazanci
%   \item A Novel Knowledge-Driven Systems Biology Approach for
% Phenotype Prediction upon Genetic Intervention
% Rui Chang, Robert Shoemaker, and Wei Wang
%   \end{itemize}
% }

\subsection{The mathematical framework }
\label{sec:cadre}

We assume that we have a gene expression dataset (e.g.\ RNA-seq data) that we will call in the following \emph{initial dataset}.
\begin{equation}
  \label{eq:def-dataset-initial}
  \mathbf{x}=(x_{i,k,p},1\leq i\leq N,1\leq k\leq K, 1\leq p\leq P),
\end{equation}
where
\begin{itemize}
\item $P$ is  the number of experiments carried out.
\item $N$ is the number of genes whose expression is measured in each experiment.
\item For each experiment, $K$ measurements are realized at times $t_1=0<
  t_2 <\ldots <t_K$.
\item $x_{i,k,p}$ is a real number which measures the expression (after normalization) of the gene $i$ at the $k$-th time of measurement
  during the $p$-th experiment. In the following, we will restrict ourselves only to absolute expression data
  $x_{i,k,p}\in\mathbb{R}_+$, to make it simpler to model the effect of gene alteration (e.g., gene knock-out reduces the absolute
  expression of the gene to 0).
\end{itemize}

Our statistical model is based on the assumption that the matrices
\[
(x_{i,k,1})_{1\leq i\leq N,1\leq k\leq K},(x_{i,k,2})_{1\leq i\leq N,1\leq k\leq K},\ldots,(x_{i,k,P})_{1\leq i\leq N,1\leq k\leq K}
\]
are realizations of independent and identically distributed (i.i.d.) experiments of the random matrix
\begin{equation}
  \label{eq:def-X}
  \mathbf{X}=(X_{i,k})_{1\leq i\leq N,1\leq k\leq K}.
\end{equation}
%dont la famille des lois possibles constitue l'ensemble des param\`etres du mod\`ele %statistique.
We suppose that the family of laws of $\mathbf{X}$ is parametrized by $\theta$ belonging to a set $\Theta$.  The statistical model is then written as :
\begin{equation}
  \label{eq:modele-stat}
  (\mathbb{P}_\theta)_{\theta\in\Theta},
\end{equation}
where $\mathbb{P}_\theta$ is the distribution of $\mathbf{X}$ when the parameter equals $\theta$. % It is assumed that the model is
% identifiable.
We will give in Section~\ref{sec:liste} several classes of models that have been considered in the literature. For example, in the Gaussian graphical model, the family $(\mathbb{P}_\theta)_{\theta\in\Theta}$ is a Gaussian family (cf. Section~\ref{sec:GGM}). A model of the type~\eqref{eq:modele-stat} will be called \emph{quantitative}.\\
% When the law of the observations is not specified, the model is said to be \emph{non-quantitative} (or  \emph{data driven}~\citep{Huynh2019survey}). We will not consider in the following such models.\\
Most models are associated with the notion of a gene network, which can be seen as a component of the model parameter. The
interpretation of the gene network may vary from one model to another. We will come back to the notion of gene network in Section~\ref{sec:liste}.\\
We define \emph{mechanistic} models as those giving the expression of each gene at time $t_k$ as an explicit function of the
expressions of all genes at the preceding times $t_1,\ldots,t_{k-1}$ and an independent random noise. In other words,
\begin{equation}
  \label{eq:modele-mecaniste}
  X_{ik}=F_{ik}\big(X_{1,[1,k-1]},X_{2,[1,k-1]},\ldots,X_{N,[1,k-1]}
% X_{1,k-1},X_{2,1},\ldots,X_{2,k-1},\ldots,X_{N,1},\ldots,X_{N,k-1}
% X_{j\ell},1\leq j\leq N,\ell<k\,
;\,U_{ik}\big),
\end{equation}
for all $i\in\{1,\ldots, N\}$ and $k\in\{2,\ldots, K\}$, where, for all $1\leq i\leq N$ and $\ell\leq K$,
\begin{equation}
  \label{eq:def-bullet}
  X_{i,[1,\ell]}=(X_{ik})_{1\leq k\leq \ell},
\end{equation}
$F_{ik}$ is a function and the random variables $U_{ik}$ are independent and are also independent from the initial condition  $(X_{i1})_{1\leq i\leq N}$.
The function $F_{ik}$ describes how the amounts of RNA expressed at times $t_1,\ldots, t_{k-1}$ influence the expression of the gene $i$ at time $t_k$. Models that cannot be written in this form will be referred to as \emph{non-mechanistic} models.\\
In a mechanistic model, the parameter $\theta$ can be defined as the vector formed by the family of
functions
$(F_{ik})_{1\leq i\leq N,1\leq k\leq K}$, the law of the vector of initial expressions  $(X_{i1})_{1\leq i\leq N}$ and the law of the
random noise  $(U_{ik})_{1\leq i\leq N,2\leq k\leq K}$. Such models are typical in causal inference~\citep{pearl2016}. Some examples are given in section~\ref{sec:liste}.

\subsection{Probabilistic approach for the prediction of gene alteration}
\label{sec:general}

We assume that the inference of the statistical model ~\eqref{eq:modele-stat} from the initial data $\mathbf{x}$ has been carried
out. We denote by $\hat{\theta}_P$ the estimator of $\theta$ obtained and we call $\mathbb{P}_{\hat{\theta}_P}$ the \emph{initial
  prediction}. Our goal is to obtain a prediction of a gene alteration from the initial prediction.

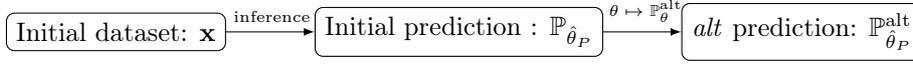
\begin{figure}
  % \captionsetup[subfigure]{justification=centering}
  % \begin{subfigure}{0.35\textwidth}
  \begin{center}
    \begin{tikzpicture}[scale=0.82]
      \node[draw,rectangle,rounded corners=3pt] (D0) at (-3,0) {\small Initial dataset: $\mathbf{x}$};
      \node[draw,rectangle,rounded corners=3pt] (M1) at (2.5,0) {\small Initial prediction : $\mathbb{P}_{\hat{\theta}_P}$};
      \node[draw,rectangle,rounded corners=3pt] (BI) at (8,0) {\small\emph{alt} prediction: $\mathbb{P}^\text{alt}_{\hat{\theta}_P}$};
      \draw[->,>=latex] (D0) -- (M1);
      \draw (-0.5,0) node [above] {\tiny inference};
      \draw[->,>=latex] (M1) -- (BI);
      \draw (5.5,0) node [above] {\tiny $\theta\mapsto\mathbb{P}^\text{alt}_\theta$};
      % \node[draw,rectangle,rounded corners=3pt] (D0) at (0,0) {Dataset 0: $\mathbf{x}$};
      % \node[draw,rectangle,rounded corners=3pt] (M1) at (0,-2) {Mod\`ele 1: $\mathbb{P}_{\hat{\theta}_P}$};
      % \node[draw,rectangle,rounded corners=3pt] (BI) at (0,-4) {Mod\`ele BI: $\mathbb{P}^{BI}_{\hat{\theta}_P}$};
      % \draw[->,>=latex] (D0) -- (M1);
      % \draw (0,-0.9) node [right] {inf\'erence};
      % \draw[->,>=latex] (M1) -- (BI);
      % \draw (0,-3) node [right] {pr\'ediction};
      % \draw (-0.2,0) node [above] {\phantom{inf\'erence}};
    \end{tikzpicture}
  \end{center}
  % \caption{Construction directe du mod\`ele BI \`a partir du mod\`ele 1}
  % \end{subfigure}
  % \begin{subfigure}{0.6\textwidth}
  %   \begin{center}
  %     \begin{tikzpicture}
  %       \node[draw,rectangle,rounded corners=3pt] (D0) at (-2.5,0) {Dataset 0: $\mathbf{x}$};
  %       \node[draw,rectangle,rounded corners=3pt] (M1) at (2.5,0) {Mod\`ele 1: $\mathbb{P}_{\hat{\theta}_P}$};
  %       \node[draw,rectangle,rounded corners=3pt] (DBI) at (0,-2) {Dataset BI: $\mathbf{x}^{BI}$};
  %       \node[draw,rectangle,rounded corners=3pt] (BI) at (0,-4) {Mod\`ele BI: $\mathbb{P}^{BI}_{\hat{\theta}'_P}$};
  %       \draw[->,>=latex] (D0) -- (M1);
  %       \draw (-0.2,0) node [above] {inf\'erence};
  %       \draw[->,>=latex] (D0) -- (DBI);
  %       \draw[->,>=latex] (M1) -- (DBI);
  %       \draw[->,>=latex] (DBI) -- (BI);
  %       \draw (0,-2.9) node [right] {inf\'erence};
  %     \end{tikzpicture}
  %   \end{center}
  %   \caption{Approche de~\citep{vallat2013reverse}: construction \`a partir du dataset 0 et du mod\`ele 1 d'un dataset BI artificiel
  %   afin d'inf\'erer le
  %   mod\`ele BI}
  % \end{subfigure}
  \caption{% Diagrammes des approches de
    Prediction of gene alteration}
  \label{fig:prediction}
\end{figure}

Our approach consists in associating with any $\theta\in\Theta$, a model $\mathbb{P}^\text{alt}_{\theta}$ describing the law of
expressions of genes after alteration. The prediction of gene alteration is then given by $\mathbb{P}^\text{alt}_{\hat{\theta}_P}$
(cf. Figure~\ref{fig:prediction}). Let us note that our approach has retained the same set of parameters for the prediction than the
initial model, but generally the model $\mathbb{P}^\text{alt}_\theta$ model does not belong to the family
$(\mathbb{P}_{\theta'})_{\theta'\in\Theta}$.

We consider two approaches to construct the laws $\mathbb{P}_\theta^\text{alt}$, one based only on the laws $(\mathbb{P}_\theta)_{\theta\in\Theta}$, the other on a mechanistic interpretation of this model.

For reasons of simplicity, we present the principle of these approaches in the case where the alteration acts only on gene 1 and
takes the form either of a gene knock-down or overexpression. We introduce a parameter $\beta\geq 0$ representing the alteration of
expression of gene 1: if $\beta\in[0,1]$, it represents the fraction of expression of gene 1 that remains after reduction of
expression; if $\beta>1$, then $\beta-1$ represents the relative overexpression of gene 1. If $\beta=0$, gene 1 is completely
knocked-out after alteration; $\beta\in(0,1)$ corresponds to a knock-down of gene 1. The case of alteration of a gene $j\neq 1$ or of
several genes simultaneously, possibly with different values of $\beta$, is easily deduced.

\subsubsection{Conditioning approach}
\label{sec:cond}

% Let us consider the case \textcolor{blue}{of a knock-out (or silencing) of gene 1}. We assume that it is effective only for a given fraction $1-\beta$
% of expressed (transcripted) RNAs. When $\beta=0$, there is no silencing, when $\beta=1$, silencing is total.\\
We denote by $\mathbb{P}_\theta^{\text{alt}, \beta}$ the prediction of silencing by conditional law. This law is
constructed by assuming that
\begin{itemize}
\item the law of  $(X_{1k})_{1\leq k\leq K}$ under  $\mathbb{P}^{\text{alt},\beta}_\theta$ is the law of $(\beta X_{1k})_{1\leq
    k\leq K}$ under $\mathbb{P}_\theta$
\item the conditional distribution of $(X_{ik})_{2\leq i\leq N,1\leq k\leq K}$ given $(X_{1k})_{1\leq k\leq K}$ is the same under
  $\mathbb{P}^{\text{alt},\beta}_\theta$ or $\mathbb{P}_\theta$.
\end{itemize}
If under the probability $\mathbb{P}_\theta$,  the vector $(X_{1k})_{1\leq k\leq K}$  has a density function $p_\theta$ on $\mathbb{R}_+^{K}$ , then for all $\beta>0$, under $\mathbb{P}^{\text{alt},\beta}_\theta$, $(X_{1k})_{1\leq k\leq K}$ has for density
\[
  p_\theta\left(\frac{x_{11}}{\beta},\ldots, \frac{x_{1K}}{\beta}\right)\frac{1}{\beta^K}
\]
%On obtient alors la loi $\mathbb{P}^{\text{BI},\beta}_\theta$ d\'efinie par
% In other words
and
\begin{multline*}
  \mathbb{P}^{\text{alt},\beta}_\theta(\mathbf{X}\in \mathrm{d} \mathbf{x}) \\ =\mathbb{P}_\theta\left(X_{2, [1,K]}\in \mathrm{d} x_{2, [1,K]},\ldots, X_{N, [1,K]}\in \mathrm{d} x_{N, [1,K]}
\mid X_{1, [1,K]}=x_{1, [1,K]}\right)  \\ \times p_\theta\left(\frac{x_{1, [1,K]}}{\beta}\right)\frac{1}{\beta^K}\mathrm{d} x_{1, [1,K]},
\end{multline*}
where $x_{i, [1,K]}$ stands for $(x_{i,1},\ldots, x_{i,K})$, $2\leq i\leq N$.\\
In the case of a knock-out of gene 1 ($\beta=0$), we obtain
\begin{multline*}
  \mathbb{P}^{\text{alt},\beta}_\theta(\mathbf{X}\in \mathrm{d} \mathbf{x}) \\ =\mathbb{P}_\theta\left(X_{2,[1,K]}\in \mathrm{d} x_{2, [1,K]},\ldots, X_{N, [1,K]}\in \mathrm{d} x_{N, [1,K]}
\mid X_{1, [1,K]}=0\right) \\ \times\delta_0(\mathrm{d} x_{1, [1,K]}),
\end{multline*}
where $\delta_0$ is the Dirac measure at $0$ in $\mathbb{R}^K$.

\subsubsection{Mechanistic approach}
\label{sec:meca}

We start from the mechanistic model~\eqref{eq:modele-mecaniste}. The mechanistic model of gene 1 alteration is based on the assumption
that the functions $F_{ik}$ and the law of the random vector $(U_{ik})_{1\leq i\leq N,1\leq k\leq K}$ are not modified by the
alteration. Moreover the vector $(U_{ik})_{1\leq i\leq N,1\leq k\leq K}$ remains independent of the initial expressions
$(X_{i1})_{1\leq i\leq N}$. This leads to the model
% Hence, the \textcolor{blue}{biological intervention model of the silencing experiment} of gene 1 is given by
\begin{equation*}
  % \label{eq:modele-meca-BI}
  \begin{cases}
    X^{\text{alt},\beta}_{11}=\beta X_{11},\quad X^{\text{alt},\beta}_{i1}=X_{i1},\ \forall i\in\{2,\ldots,N\}, & \\
    X^{\text{alt},\beta}_{ik}=F_{ik}\left(\beta X^{\text{alt},\beta}_{1, [1,k-1]},X^{\text{alt},\beta}_{2, [1,k-1]},\ldots,
      X^{\text{alt},\beta}_{N, [1,k-1]};\,U_{ik}\right), &
  \end{cases}
\end{equation*}
for all  $1\leq i\leq N$ and $2\leq k\leq K$.
% , where the joint distribution of initial data $(X^{\text{alt},\beta}_{11},\ldots,X^{\text{alt},\beta}_{N1})$ and
% $(U^{\text{alt},\beta}_{ik})_{2\leq i\leq N,2\leq k\leq K}$ is the same as that of $(X_{11},\ldots,X_{N1})$ and
% $(U_{ik})_{2\leq i\leq N,2\leq k\leq K}$ under $\mathbb{P}_\theta$.
The probability $\mathbb{P}^{\text{alt},\beta}_\theta$  is the law of the random vector
$\mathbf{X}^{\text{alt},\beta}$.\\
% The model of
% silencing for the quantity of active RNA is easily deduced from the relation~\eqref{eq:translated-transcript}.
Note that the
previous gene alteration model is also mechanistic. In the case of a knock-out of gene 1 ($\beta=0$), we obtain
\begin{equation}
  \label{eq:modele-meca-BI}
  X^{\text{alt},0}_{ik}=F_{ik}\left(0,X^{\text{alt},0}_{2, [1,k-1]},\ldots,X^{\text{alt},0}_{N, [1,k-1]}
% X_{1,k-1},X_{2,1},\ldots,X_{2,k-1},\ldots,X_{N,1},\ldots,X_{N,k-1}
% X_{j\ell},1\leq j\leq N,\ell<k\,
;\,U_{ik}\right),
\end{equation}
for all $i\in\{2,\ldots, N\}$ and $k\in\{2,\ldots, K\}$.% \\
% We could similarly consider the case of partial silencing of one gene $j\neq 1$ or the case of simultaneous partial silencing of several genes $j\neq 1$.
% or the case of simultaneous partial silencing of several genes.

\subsubsection{A simple example}
\label{sec:exemple}

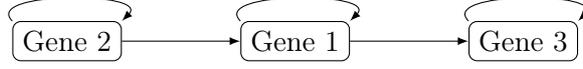
\begin{figure}[h]
  \begin{center}
    \begin{tikzpicture}
      \node[draw,rectangle,rounded corners=3pt] (G2) at (0,0) {Gene 2};
      \node[draw,rectangle,rounded corners=3pt] (G1) at (3,0) {Gene 1};
      \node[draw,rectangle,rounded corners=3pt] (G3) at (6,0) {Gene 3};
      \draw[->,>=latex] (G2) -- (G1);
      \draw[->,>=latex] (G1) -- (G3);
      \draw[->,>=latex] (G1.north west) to[out=135,in=45] (G1.north east);
      \draw[->,>=latex] (G2.north west) to[out=135,in=45] (G2.north east);
      \draw[->,>=latex] (G3.north west) to[out=135,in=45] (G3.north east);
    \end{tikzpicture}
  \end{center}
  \caption{Dependency network between the expressions of genes 1, 2 and 3}
  \label{fig:ex-simple}
\end{figure}

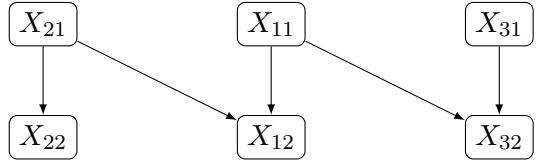
\begin{figure}[h]
  \begin{center}
    \begin{tikzpicture}
      \node[draw,rectangle,rounded corners=3pt] (X21) at (0,1.5) {$X_{21}$};
      \node[draw,rectangle,rounded corners=3pt] (X11) at (3,1.5) {$X_{11}$};
      \node[draw,rectangle,rounded corners=3pt] (X31) at (6,1.5) {$X_{31}$};
      \node[draw,rectangle,rounded corners=3pt] (X22) at (0,0) {$X_{22}$};
      \node[draw,rectangle,rounded corners=3pt] (X12) at (3,0) {$X_{12}$};
      \node[draw,rectangle,rounded corners=3pt] (X32) at (6,0) {$X_{32}$};
      \draw[->,>=latex] (X21) -- (X22);
      \draw[->,>=latex] (X21) -- (X12);
      \draw[->,>=latex] (X11) -- (X12);
      \draw[->,>=latex] (X11) -- (X32);
      \draw[->,>=latex] (X31) -- (X32);
    \end{tikzpicture}
  \end{center}
  \caption{Bayesian dependency network of  $(X_{ik})_{1\leq i\leq 3,\,1\leq k\leq 2}$ under $\mathbb{P}_\theta$.}
  \label{fig:ex-reseau-bayes}
\end{figure}

To illustrate the difference between the mechanistic and conditioning approaches, let us consider an example involving 3 genes and
two dates ($N=3$, $K=2$), with dependency relationships between genes shown in Figure~\ref{fig:ex-simple} (compact representation
without time-dependence) and Figure~\ref{fig:ex-reseau-bayes} (full Bayesian dependency network).
%Ce graphique signifie que les expressions de g\`enes %$(X_{ik})_{1\leq i\leq 3,\,1\leq k\leq 2}$ satisfont le %r\'eseau de d\'ependance  de %% la %%Figure~\ref{fig:ex-reseau-bayes}.}

Let us  consider the knock-out of gene 1 ($\beta=0$). For simplicity, we assume that the random variables
 $X_{ik}$ take values in a discrete set $\mathcal{X}$ containing 0. In the formalism of Bayesian networks (cf.\ section~\ref{sec:bayes}), Figure~\ref{fig:ex-reseau-bayes}  means that for all
 $\theta\in\Theta$, the law of the random vector
$(X_{ik})_{1\leq i\leq 3,\,1\leq k\leq 2}$ is
\begin{align*}
  & \mathbb{P}_\theta(X_{11}=x_{11},X_{21}=x_{21},X_{31}=x_{31},X_{12}=x_{12},X_{22}=x_{22},X_{32}=x_{32}) \\
  & =\mathbb{P}_\theta(X_{11}=x_{11}) \mathbb{P}_\theta(X_{21}=x_{21})
  \mathbb{P}_\theta(X_{31}=x_{31})\mathbb{P}_\theta(X_{22}=x_{22}\mid X_{21}=x_{21}) \\ & \qquad\times\mathbb{P}_\theta(X_{12}=x_{12}\mid
  X_{11}=x_{11},X_{21}=x_{21}) \\ & \qquad\times\mathbb{P}_\theta(X_{32}=x_{32}\mid X_{11}=x_{11},X_{31}=x_{31}),
\end{align*}
for all $(x_{ik})_{1\leq i\leq 3,\,1\leq k\leq 2}$ in $\mathcal{X}^6$. We suppose that $\mathbb{P}_\theta(X_{11}=0,\,X_{12}=0)>0$.
% \\ Our objective is to show that our two approaches to model \textcolor{blue}{silencing} produce different results.

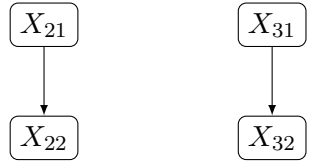
\begin{figure}[h]
  \begin{center}
    \begin{tikzpicture}
      \node[draw,rectangle,rounded corners=3pt] (X21) at (0,1.5) {$X_{21}$};
      \node[draw,rectangle,rounded corners=3pt] (X31) at (3,1.5) {$X_{31}$};
      \node[draw,rectangle,rounded corners=3pt] (X22) at (0,0) {$X_{22}$};
      \node[draw,rectangle,rounded corners=3pt] (X32) at (3,0) {$X_{32}$};
      \draw[->,>=latex] (X21) -- (X22);
      \draw[->,>=latex] (X31) -- (X32);
    \end{tikzpicture}
  \end{center}
  \caption{Bayesian dependency network for  $(X_{ik})_{1\leq i\leq 3,\,1\leq k\leq 2}$ under  $\mathbb{P}^{\text{alt,cond}}_\theta$.}
  \label{fig:ex-reseau-cond}
\end{figure}

\paragraph{Conditioning approach}
Let us denote by $\mathbb{P}^{\text{alt,cond}}_\theta$ the law $\mathbb{P}^{\text{alt},0}_\theta$ obtained using the conditioning approach. By definition:
\begin{multline*}
  \mathbb{P}^{\text{alt,cond}}_\theta(X_{21}=x_{21},X_{31}=x_{31},X_{22}=x_{22},X_{32}=x_{32}) \\
  =\mathbb{P}_\theta(X_{21}=x_{21},X_{31}=x_{31},X_{22}=x_{22},X_{32}=x_{32}\mid
  X_{11}=X_{12}=0) % \\ & =\mathbb{P}_\theta(X_{31}=x_{31}) \mathbb{P}_\theta(X_{32}=x_{32}\mid X_{11}=0,X_{31}=x_{31}) \\ &
  % \qquad\times\frac{\mathbb{P}_\theta(X_{21}=x_{21}) \mathbb{P}_\theta(X_{12}=0\mid
  % X_{11}=0,X_{21}=x_{21})}{\sum_{y_{21}}\mathbb{P}_\theta(X_{21}=y_{21})\mathbb{P}_\theta(X_{12}=0\mid X_{11}=0,X_{21}=y_{21})}
  % \\ & \qquad\times\mathbb{P}_\theta(X_{22}=x_{22}\mid X_{21}=x_{21}).
  % \end{aligned}
\end{multline*}
Using the independence of the random variables $X_{11}$ $X_{21}$ and $X_{31}$ we obtain after some computation
\begin{multline}
  \mathbb{P}^{\text{alt,cond}}_\theta(X_{21}=x_{21},X_{31}=x_{31},X_{22}=x_{22},X_{32}=x_{32}) \\
    =\mathbb{P}_\theta(X_{31}=x_{31}) \mathbb{P}_\theta(X_{32}=x_{32}\mid X_{11}=0,X_{31}=x_{31}) \\
    \times \mathbb{P}_{\theta}(X_{21}=x_{21}\mid X_{11}=X_{12}=0)\mathbb{P}_\theta(X_{22}=x_{22}\mid X_{21}=x_{21}).
    \label{eq:ex-cond}
\end{multline}
We derive the following results
\begin{align}
  \mathbb{P}^{\text{alt,cond}}_\theta(X_{31}=x_{31}) & =\mathbb{P}_\theta(X_{31}=x_{31}), \notag \\
  \mathbb{P}^{\text{alt,cond}}_\theta(X_{32}=x_{32}\mid X_{31}=x_{31}) & =\mathbb{P}_\theta(X_{32}=x_{32}\mid X_{11}=0,X_{31}=x_{31}),
  \notag \\
  \mathbb{P}^{\text{alt,cond}}_\theta(X_{21}=x_{21}) & =\mathbb{P}_\theta(X_{21}=x_{21}\mid X_{12}=X_{11}=0), \label{eq:ex-fin} \\
  \mathbb{P}^{\text{alt,cond}}_\theta(X_{22}=x_{22}\mid X_{21}=x_{21}) & =\mathbb{P}_\theta(X_{22}=x_{22}\mid X_{21}=x_{21}). \notag
\end{align}
In other words, the Bayesian dependency network associated with the model $\mathbb{P}^{\text{alt,cond}}_\theta$ is given by
Figure~\ref{fig:ex-reseau-cond}. This network is obtained by deleting from the network in Figure~\ref{fig:ex-reseau-bayes},  all edges originating from or directed towards one of the two vertices
$X_{11}$ and $X_{12}$. In the sequel, we call the edges \emph{incident edges to $X_{11}$ and to $X_{12}$.}

\paragraph{Mechanistic approach}
Let us denote by $\mathbb{P}^{\text{alt,meca}}_\theta$ the law $\mathbb{P}^{\text{alt},0}_\theta$ obtained using the mechanistic
approach. \\
The mechanistic approach requires the model to be written as~\eqref{eq:modele-mecaniste}. Consider the random variables
$(X^\text{meca}_{11},X^\text{meca}_{21},X^\text{meca}_{31})=(X_{11},X_{21},X_{31})$ under $\mathbb{P}_\theta$. Let
$U_{12},U_{22},U_{32}$ i.i.d. random variables uniformly distributed on  $[0,1]$ and independent of
$(X^\text{meca}_{11},X^\text{meca}_{21},X^\text{meca}_{31})$.  For any $x_{11},x_{21},x_{31}$, we introduce the functions:
\begin{itemize}
\item $F_{22}(x_{21};\cdot)$ is the generalized inverse of the distribution function of $X_{22}$ under $\mathbb{P}_{\theta}$ conditional on
  $X_{21}=x_{21}$;
\item $F_{12}(x_{11},x_{21};\cdot)$ is the generalized inverse of the distribution function of  $X_{12}$ conditional on $X_{11}=x_{11}$ and
  $X_{21}=x_{21}$;
\item $F_{32}(x_{11},x_{31};\cdot)$ is the generalized inverse of the distribution function of  $X_{32}$ conditional on  $X_{11}=x_{11}$ and
  $X_{31}=x_{31}$.
\end{itemize}
We now define the vector
 $(X^\text{meca}_{ik},1\leq i\leq 3,1\leq k\leq 2)$ as follows
\[
  X^\text{meca}_{22}=F_{22}\left(X^\text{meca}_{21};U_{22}\right),\quad X^\text{meca}_{12}=F_{12}\left(X^\text{meca}_{11},X^\text{meca}_{21};U_{12}\right)
\]
and
\[
  X^\text{meca}_{32}=F_{32}\left(X^\text{meca}_{11},X^\text{meca}_{31};U_{32}\right).
\]
Then $(X^\text{meca}_{ik},1\leq i\leq 3,1\leq k\leq 2)$ has the same distribution as  $(X_{ik},1\leq i\leq 3,1\leq k\leq 2)$ under $\mathbb{P}_\theta$.\\
According to~\eqref{eq:modele-meca-BI}, the mechanistic approach to model the knock-out of gene 1 leads to the model
\begin{gather*}
  \left(X^{\text{alt,meca}}_{21},X^{\text{alt,meca}}_{31}\right)=\left(X^\text{meca}_{21},X^\text{meca}_{31}\right), \\
  X^{\text{alt,meca}}_{22}=F_{22}\left(X^{\text{alt,meca}}_{21};
  U_{22}\right)\text{ and } X^{\text{alt,meca}}_{32}=F_{32}\left(\,0\,,X^{\text{alt,meca}}_{31};
  U_{32}\right).
\end{gather*}
From the definition of functions $F_{22}$ and $F_{32}$, we deduce that
\begin{multline*}
  \mathbb{P}^{\text{alt,meca}}_\theta\left(X^{\text{alt,meca}}_{21}=x_{21},
  X^{\text{alt,meca}}_{31}=x_{31},X^{\text{alt,meca}}_{22}=x_{22},
  X^{\text{alt,meca}}_{32}=x_{32}\right) \\ =\mathbb{P}_\theta(X_{21}=x_{21})
  \mathbb{P}_\theta(X_{31}=x_{31}) \mathbb{P}_\theta(X_{22}=x_{22}\mid X_{21}=x_{21}) \\ \mathbb{P}_\theta(X_{32}=x_{32}\mid X_{11}=0,X_{31}=x_{31}).
\end{multline*}
From this we deduce
\[
\mathbb{P}^{\text{alt,meca}}_\theta\left(X^{\text{alt,meca}}_{21}=x_{21}\right)=\mathbb{P}_\theta(X_{21}=x_{21}).
\]

A comparison with~\eqref{eq:ex-fin} shows that the two models are not identical. This difference can be explained by the fact that
conditioning the expression of gene 2 at time 2 on the fact that the expressions of gene 1 are zero is a different constraint from
that of the mechanistic model: conditioning has the effect of biasing the expression of genes located upstream of gene 1. More will
be said on the differences between the two models in Section~\ref{sec:bayes}.

\section{Applications of the gene alteration modeling}
\label{sec:utilite}

We consider two distinct settings for the application of our method. First, on the basis of the initial dataset only, we consider the
problem of design of experiments, i.e.\ the choice of gene alteration experiments among several possibilities, in order to best
achieve a given objective (Section~\ref{sec:choix-xp}). Second, assuming that gene alteration data were obtained in addition to the
initial dataset, we can define a statistical model of these two datasets, allowing to address statistical issues such as validation
or estimation (Section~\ref{sec:stat}).
% We give three examples of how these models can be used. The first concerns the choice \textcolor{blue}{of a silencing experiment to
% best achieve a given objective. When a set of silenced data is available, the second application aims at validating certain
% elements of the biological intervention model,} and the third aims at improving the estimation of the parameters of the initial
% model.
We will apply these methods to typical models of gene regulatory networks in Section~\ref{sec:liste}.

\subsection{Application for the design of experiments}
\label{sec:choix-xp}

Let us consider the case where several possible gene alteration experiments $\text{alt}_1,\ldots,\text{alt}_n$ are possible and the
objective is to favor a certain cell behavior such as proliferation~\citep{Schleiss2021}, or more generaly maximize some
utility~\citep{Venkat2017,Bouvier2024}. In a biological context different from gene regulatory networks, maximizations of measures of
vulnerability~\citep{Zhu2016} or robustness~\citep{Bianconi2015} were also considered using methodological approaches. In the
sequel, we consider the objective of reducing proliferation.\\
The search for a predictor of cellular behavior such as proliferation is a difficult task that can be address using various
methods~\citep{Azuaje2010,Bianconi2015,Zhu2016,Venkat2017}, e.g.\ using a learning method or known biological information on the genes
involved in the cellular proliferation program. We assume here that we have a model of (possibly nonlinear) regression of the
response variable of proliferation $Y$ in terms of the explanatory variables given by the gene expression matrix $\mathbf{X}$ as
follows
\[
  Y=H(\mathbf{X}),
\]
where $H$ can be random. \\
We then choose the experiment $i$ that minimizes the mean proliferation scores
\[
  \mathbb{E}_{\mathbb{P}^{\text{alt}_1}_{\hat{\theta}_P}} H(\mathbf{X}),\ldots, \mathbb{E}_{\mathbb{P}^{\text{alt}_n}_{\hat{\theta}_P}} H(\mathbf{X}).
\]
A simple example of predictor $H$ is given by
\[
  H(\mathbf{X})=\sum_{i\in G}\sum_{k\in T}X_{i,k},
\]
where $G\subset\{1,\ldots,N\}$ is a set of genes involved in the proliferation program and $T\subset\{1,\ldots,K\}$ is the set of
relevant dates.\\
% We can then choose
% In this case, we select the silencing experiment $l$ that maximizes
% \[
%   \sum_{i\in G}\sum_{k\in T}\mathbb{E}_{\mathbb{P}^{\text{BI}_\ell}_{\hat{\theta}_P}} X_{i,k},\quad 1\leq \ell\leq n.
% \]
% {\color{magenta} The predictor of proliferation proposed in~\citep{these-Rodolphe} fits into this formalism [Laurent, est-ce qu'on en parle ?]}.
Another method of selection of gene alteration experiments would be to choose the experiment that maximizes the number of
down-regulated genes in $G$ at the dates in $T$. The gene $i$ is said to be up-regulated (resp. down-regulated) at time $k$ for
the experiment $\text{alt}_\ell$ if
\[
  \mathbb{E}_{\hat{\theta}_P} X_{i,k}<\mathbb{E}_{\mathbb{P}^{\text{alt}_\ell}_{\hat{\theta}_P}} X_{i,k} \quad \left(\text{resp.}
  \quad \mathbb{E}_{\hat{\theta}_P} X_{i,k}>\mathbb{E}_{\mathbb{P}^{\text{alt}_\ell}_{\hat{\theta}_P}} X_{i,k}\right).
\]
We then define the predictor of proliferation associated with experiment $\text{alt}_\ell$ as
\[
  S_\ell=\text{Card}\left\{(i,k)\in G\times T: \mathbb{E}_{\hat{\theta}_P} X_{i,k}>\mathbb{E}_{\mathbb{P}^{\text{alt}_\ell}_{\hat{\theta}_P}} X_{i,k}\right\},
\]
where $\text{Card(A)}$ is the number of elements in the set $A$, and we select the gene alteration experiment that maximizes $S_1,\ldots,S_n$.\\
Another possibility is to choose the alteration experiment that most closely approximates gene expressions to given mean target values
$m_{ik}^\text{target}$ (e.g. measured in non proliferative cells). In this case
\[
  S_\ell=\sum_{i\in G}\sum_{k\in T}\left|\mathbb{E}_{\mathbb{P}^{\text{alt}_\ell}_{\hat{\theta}_P}} X_{i,k}-m_{ik}^\text{target}\right|
\]
and we select the alteration experiment that minimizes $S_1,\ldots, S_n$.

\subsection{Application to the statistical modeling of initial and gene alteration datasets}
\label{sec:stat}
%Implications pour la validation

We now assume that we have a second dataset from a gene alteration experiment with parameter $\beta$
\[
\mathbf{x}^\text{alt}=(x^\text{alt}_{i,k,p},1\leq i\leq N,1\leq k\leq K, 1\leq p\leq P^\text{alt}),
\]
called \emph{alt dataset}. Using one of the approaches described in Section~\ref{sec:general}, we build a statistical model
% $(\mathbb{P}_\theta,\mathbb{P}^{\text{alt},\beta}_\theta)_{\theta\in\Theta}$
of the initial and alt datasets under which the likelihood of the data writes in the discrete case as
\[
\prod_{p=1}^P\mathbb{P}_\theta(\mathbf{X}=\mathbf{x}_{p})\prod_{q=1}^{P^\text{alt}}
\mathbb{P}^{\text{alt},\beta}_\theta(\mathbf{X}^{\text{alt}}=\mathbf{x}^\text{alt}_q),
\]
where $\mathbf{x}_p=(x_{i,k,p})_{1\leq i\leq N,1\leq k\leq K}$ and  $\mathbf{x}^\text{alt}_q=(x^\text{alt}_{i,k,q})_{1\leq i\leq N,1\leq k\leq K}$.

We can then use this statistical model to answer statistical questions. In the sequel, we focus on the validation of the \emph{alt} model
(Section~\ref{sec:validation}) and on the improvement of estimation of the initial model using alt data (Section~\ref{sec:affinage}).
Recall that $\hat{\theta}_P$ is an estimator of $\theta$ obtained from the \emph{initial dataset}.

\subsubsection{Validation of a model of gene alteration}
\label{sec:validation}

\begin{figure}
  \captionsetup[subfigure]{justification=centering}
  % \begin{subfigure}{1.0\textwidth}
    \begin{center}
      \begin{tikzpicture}
        \node[draw,rectangle,rounded corners=3pt] (D0) at (-0.5,0) {initial dataset: $\mathbf{x}$};
        \node[draw,rectangle,rounded corners=3pt] (M1) at (6.4,0) {initial model: $\mathbb{P}_{\hat{\theta}_P}$};
        \node[draw,rectangle,rounded corners=3pt] (BI) at (6.4,-2) {\emph{alt} model prediction: $\mathbb{P}^{\text{alt},\beta}_{\hat{\theta}_P}$};
        \node[draw,rectangle,rounded corners=3pt] (DBI) at (-0.5,-2) {\emph{alt} dataset: $\mathbf{x}^\text{alt}$};
        \draw[->,>=latex] (D0) -- (M1);
        \draw (2.85,0) node [above] {inference};
        \draw[->,>=latex] (M1) -- (BI);
        % \draw (4.4,-0.9) node [right] {pr\'ediction};
        \draw[<->,>=latex,dashed] (BI) -- (DBI);
        \draw (2.6,-2) node [above] {\small goodness-of-fit test};
      \end{tikzpicture}
    \end{center}
    \caption{Gene alteration model validation}
  \label{fig:validation}
\end{figure}
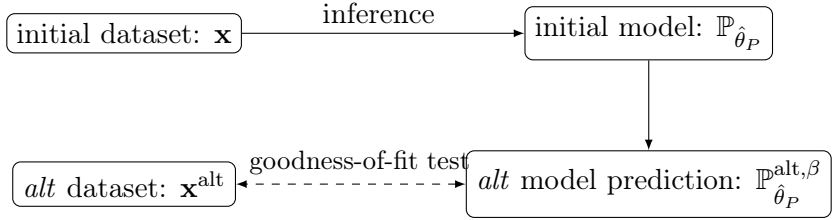

% Suppose that in addition to the \emph{initial dataset}, we also have a
% second set of data from a \textcolor{blue}{ biological intervention experiment}
% \[
% \mathbf{x}^\text{alt}=(x^\text{alt}_{i,k,p},1\leq i\leq N,1\leq k\leq K, 1\leq p\leq P^\text{alt}),
% \]
% called \emph{alt dataset}. \\
% In the following, we choose a biological intervention model $\mathbb{P}^\text{alt}_{\hat{\theta}_P}$, from the two defined in Sections
% ~\ref{sec:cond} and~\ref{sec:meca}.\\
Our aim is to check if the gene alteration model $\mathbb{P}^{\text{alt},\beta}_{\hat{\theta}_P}$ correctly accounts for the
observations $\mathbf{x}^\text{alt}$. From a statistical point of view, this corresponds to a goodness-of-fit test
(Figure~\ref{fig:validation}). Due to the small amount of observation, it is usually impossible in practice to test if
$\mathbf{x}^\text{alt}$ follows the distribution $\mathbb{P}^{\text{alt},\beta}_{\hat{\theta}_P}$. This explains why authors usually
restrict to low-dimensional test statistics relevant for the biological application, typically either relating to gene
expressions~\citep{Vallat2013}, or to the presence or absence of edges in the
network~\citep{Wagner2001,Markowetz2005}.

% Note that the chosen goodness-of-fit test is neither described nor explicitely
% performed in any of the previous references: at best, a $p$-value is calculated. This is why we describe in detail below how the test
% should be performed. We will focus on statistics constructed from average gene expressions of the alt dataset.

Below we describe the goodness-of-fit test used in~\citep{Vallat2013}, which can be adapted to any quantitative model. The idea is to
test whether the number of up- and down-regulated genes correctly predicted by the model
$\mathbb{P}^{\text{alt},\beta}_{\hat{\theta}_P}$ is significantly greater than what would be obtained if up- and down-regulated genes
were chosen at random. We consider an alteration of gene 1 only, with parameter $\beta$.
% Let $X^\text{alt}$ be a random variable with values in $N\times K$-dimensional real matrices.

More precisely, for any positive $\epsilon$, we define a partition of gene-date pairs based on the two probabilities
$\mathbb{P}_{\hat{\theta}_P}$ and $\mathbb{P}^{\text{alt},\beta}_{\hat{\theta}_P}$ as follows:
\begin{align}
  \mathcal{U} & :=\left\{
    (i,k)\in\{2,\ldots,N\}\times\{2,\ldots,K\}:\mathbb{E}_{\mathbb{P}^{\text{alt},\beta}_{\hat{\theta}_P}}(X_{i,k})\geq (1+\varepsilon)\mathbb{E}_{\mathbb{P}_{\hat{\theta}_P}}(X_{i,k})
  \right\}, \notag \\
  \mathcal{D} & :=\left\{
    (i,k)\in\{2,\ldots,N\}\times\{2,\ldots,K\}:\mathbb{E}_{\mathbb{P}^{\text{alt},\beta}_{\hat{\theta}_P}}(X_{i,k})\leq(1-\varepsilon)\mathbb{E}_{\mathbb{P}_{\hat{\theta}_P}}(X_{i,k})
  \right\}, \notag \\
  \mathcal{C} & :=\Big\{
    (i,k)\in\{2,\ldots,N\}\times\{2,\ldots,K\}: \\ & \qquad\qquad\qquad\qquad \left.(1-\varepsilon)\mathbb{E}_{\mathbb{P}_{\hat{\theta}_P}}(X_{i,k})<\mathbb{E}_{\mathbb{P}^{\text{alt},\beta}_{\hat{\theta}_P}}(X_{i,k})<(1+\varepsilon)\mathbb{E}_{\mathbb{P}_{\hat{\theta}_P}}(X_{i,k})
  \right\}. \label{eq:def-UDC}
\end{align}
The set $\mathcal{U}$ (resp.\ $\mathcal{D}$, resp.\ $\mathcal{C}$) is the set of \emph{up-regulated} (resp. \emph{down-regulated},
\emph{constant}) gene-date pairs, i.e. those whose average expression is higher in the \emph{alt} model than in the initial model at
the date considered. The sets $\mathcal{U}$, $\mathcal{D}$ and $\mathcal{C}$ contain neither gene 1 nor the first date, because gene
expression is identical in both models at this date. % Obviously, $\mathcal{C}=\emptyset$ if $\varepsilon=0$.
Note that these sets are
constructed solely on the basis of the estimated parameter
$\hat{\theta}_P$, considered here as a constant.\\
Then, we consider  the following test statistic constructed from a \emph{alt} sample $\mathbf{X}^\text{alt}=(X^\text{alt}_1,\ldots,X^\text{alt}_{P^\text{alt}})$ of size $P^\text{alt}$:
\begin{equation}
  \label{eq:statistique-S}
  % S=\frac{\#\left\{(i,k)\in\mathcal{U}:\bar{X}^\text{BI}_{i,k}>\mathbb{E}_{\mathbb{P}_{\hat{\theta}_P}}(X_{i,k})\right\}
  %   +\#\left\{(i,k)\in\mathcal{D}:\bar{X}^\text{BI}_{i,k}\leq\mathbb{E}_{\mathbb{P}_{\hat{\theta}_P}}(X_{i,k}) \right\}}{(N-1)(K-1)},
S=S(\mathbf{X}^\text{alt}):=\frac{\text{Card}(\mathcal{U}\cap\hat{\cal U}^\text{alt})+\text{Card}(\mathcal{D}\cap\hat{\cal
      D}^\text{alt})+\text{Card}(\mathcal{C}\cap\hat{\cal C}^\text{alt})}{(N-1)(K-1)},
\end{equation}
where
\begin{align*}
  \hat{\mathcal{U}}^\text{alt} & :=\left\{
    (i,k)\in\{2,\ldots,N\}\times\{2,\ldots,K\}: \bar{X}^\text{alt}_{i,k}\geq (1+\varepsilon)\mathbb{E}_{\mathbb{P}_{\hat{\theta}_P}}(X_{i,k})
  \right\}, \\
  \hat{\mathcal{D}}^\text{alt} & :=\left\{
    (i,k)\in\{2,\ldots,N\}\times\{2,\ldots,K\}: \bar{X}^\text{alt}_{i,k}\leq(1-\varepsilon)\mathbb{E}_{\mathbb{P}_{\hat{\theta}_P}}(X_{i,k})
  \right\}, \\
  \hat{\mathcal{C}}^\text{alt} & :=\Big\{
    (i,k)\in\{2,\ldots,N\}\times\{2,\ldots,K\}: \\ & \qquad\qquad\qquad\qquad
                                                     \left.(1-\varepsilon)\mathbb{E}_{\mathbb{P}_{\hat{\theta}_P}}(X_{i,k})<
                                                     \bar{X}^\text{alt}_{i,k}<(1+\varepsilon)\mathbb{E}_{\mathbb{P}_{\hat{\theta}_P}}(X_{i,k})
  \right\},
\end{align*}
with
\[
\bar{X}^\text{alt}_{i,k}:=\frac{1}{P^\text{alt}}\sum_{p=1}^{P^\text{alt}} (X^\text{alt}_p)_{i,k}.
\]
The $S$ statistic counts the proportion of mean up-regulated, down-regulated or constant gene-date pairs in the \emph{alt} sample
that agree with the model.

We consider the null and alternative hypotheses
\begin{description}
\item[\textmd{(H0)}] the distribution of $X^{\text{alt}}$ is $\mathbb{P}^{\text{alt},\beta}_{\hat{\theta}_P}$,
\item[\textmd{(H1)}] the law of $S(\mathbf{X}^\text{alt})$ is $\mu_0$,
\end{description}
where $\mu_0$ is the distribution of
\[
  S':=\frac{\text{Card}(\mathcal{U}\cap{\cal U}')+\text{Card}(\mathcal{D}\cap{\cal
      D}')+\text{Card}(\mathcal{C}\cap{\cal C}')}{(N-1)(K-1)},
\]
where
\begin{align*}
  \mathcal{U}' & :=\left\{
    (i,k)\in\{2,\ldots,N\}\times\{2,\ldots,K\}: Z_{i,k}=u\right\}, \\
  \mathcal{D}' & :=\left\{
    (i,k)\in\{2,\ldots,N\}\times\{2,\ldots,K\}: Z_{i,k}=d\right\}, \\
  \mathcal{C}' & :=\left\{
    (i,k)\in\{2,\ldots,N\}\times\{2,\ldots,K\}: Z_{i,k}=c\right\}
\end{align*}
and $\big(Z_{i,k}, (i,k)\in\{2,\ldots,N\}\times\{2,\ldots,K\}\big)$ is a sequence of i.i.d.\ random variable with values in
$\{u,d,c\}$ and distribution $(p_u,p_d,p_c)$ where
%Dans~\citep{Vallat2013}, les auteurs consid\`erent la loi $\mu_0$ de %$S$ que l'on obtiendrait en tirant ind\'ependamment les
%g\`enes up-r\'egul\'es, down-r\'egul\'es et constants avec des %probabilit\'es
\[
  p_u:=\frac{\text{Card}(\mathcal{U})}{(N-1)(K-1)},\quad
  p_d:=\frac{\text{Card}(\mathcal{D})}{(N-1)(K-1)},\quad p_c:=1-p_u-p_d.
\]
Note that $S'$ is obtained from $S$ by replacing $\hat{\mathcal{U}}^\text{alt}$,
$\hat{\mathcal{D}}^\text{alt}$ and $\hat{\mathcal{C}}^\text{alt}$ with ``purely'' random sets $\mathcal{U}'$, $\mathcal{D}'$ and
$\mathcal{C}'$. In particular, the random variables $\text{Card}(\mathcal{U}\cap{\cal U}')$,
$\text{Card}(\mathcal{D}\cap{\cal D}')$ and  $\text{Card}(\mathcal{C}\cap{\cal C}')$ are independent, with binomial laws of
parameters $(\text{Card}(\mathcal{U}),p_u)$, $(\text{Card}(\mathcal{D}),p_d)$ and $(\text{Card}(\mathcal{C}),p_c)$, respectively.

Hypothesis (H0) will be rejected for low values of the statistic $S$. Hence the critical region associated with the test of
significance level $\alpha$ (typically, $\alpha=5 \%$) is an interval $R=[0,u_\alpha]$, where $u_\alpha>0$ is determined by the condition
\begin{equation}
  \label{eq:region-rejet}
  \mathbb{P}_{\hat{\theta}_p}^{\text{alt},\beta}\left( S(\mathbf{X}^\text{alt})\in R\right)\leq\alpha,
\end{equation}
If $\hat{s}:=S(\mathbf{x}^\text{alt})$ is smaller than $u_\alpha$, (H0) is rejected. Otherwise, the $p$-value of the test is
$\mathbb{P}_{\hat{\theta}_p}^{\text{alt},\beta}\left( S(\mathbf{X}^\text{alt})\leq
  \hat{s}\right)$ and the power of the test is given by the type II error $\mu_0(R^c)$,
which will be small if the sample size is large enough.

\subsubsection{Improved estimation of initial model parameters}
\label{sec:affinage}

%Implications pour l'am\'elioration d'estimation des pa\-ra\-m\`e\-tres d'un mod\`ele
% As in the previous section, let us consider the case where we have an initial dataset \textbf{x} and an alt dataset $\mathbf{x}^\text{alt}$
%  derived from a biological intervention experiment and independent of \textbf{x}.\\
We now want to understand how the $\mathbf{x}^\text{alt}$ dataset can be used to improve the estimation of the parameter $\theta$ of
the initial model $(\mathbb{P}_\theta)_{\theta\in\Theta}$. We consider here the usual cases where parameter inference is based either
on likelihood maximization or on a posteriori likelihood maximization methods.\\
Assume that for any $\theta\in\Theta$, the probability measures $\mathbb{P}_\theta$ and $\mathbb{P}^{\text{alt},\beta}_\theta$ on
$\mathbb{R}^{NK}$ are absolutely continuous with respect to a reference measure $\rho$. Let $f_\theta$ and
$g^{\text{alt},\beta}_\theta$ denote respectively the densities of these laws. In the case of a posteriori maximization, we assume
that the a priori law has a density $p(\theta)$ with respect to a reference measure on $\Theta$. In the case of likelihood
maximization, let $p(\theta)=1$, for any
$\theta\in\Theta$.\\
The likelihood of the data $(\mathbf{x},\mathbf{x}^\text{alt})$ is then
\begin{equation}
  \label{eq:amelioration-estimation}
  p(\theta)\prod_{p=1}^P f_\theta(\mathbf{x}_p)\prod_{q=1}^{P^\text{alt}} g^{\text{alt},\beta}_\theta(\mathbf{x}^\text{alt}_q).
\end{equation}
% where $\mathbf{x}_p=(x_{i,k,p})_{1\leq i\leq N,1\leq k\leq K}$ and  $\mathbf{x}^\text{alt}_q=(x^\text{alt}_{i,k,q})_{1\leq i\leq N,1\leq k\leq K}$.
The new estimate of parameter $\theta$ is obtained by maximizing the previous quantity with respect to $\theta\in\Theta$ (see the examples of Section~\ref{sec:liste}).
% \\
% We will study in greater detail an example of maximum a posteriori estimation in the case of the LiRE model in section \ref{sec:lasso}.

\section{Gene alteration modeling for quantitative and mechanistic models}
\label{sec:liste}

% Our goal is to implement the methods \textcolor{blue}{for predicting a biological intervention experiment} that have been
% described in section~\ref{sec:construction-BI} and examine how the resulting models can be used to validate or improve
% estimates using the approaches given in section ~\ref{sec:utilite}.

% \textcolor{blue}{j'ai supprim\'e la suite: mais pourraient fournir une alternative
% int\'eressante pour la pr\'ediction du silencing.}

In the following, we will consider Gaussian graphical models (which are non-mechanistic, but quantitative models) and two mechanistic
models: Bayesian networks and penalized linear regressions. Our goal is to implement for each model the methods of
Sections~\ref{sec:construction-IB} and~\ref{sec:utilite}.

% In each case, we will start with a short description of the
% model and then apply our alt scheme.

\subsection{Gaussian graphical models}
\label{sec:GGM}

% Gaussian graphical models offer no natural mechanistic interpretation.
There are many variants of these models but we will
concentrate on the simplest, for which we show below that the model of gene knock-out is explicit and its
impact on the gene regulatory network can be fully characterized. As an application, we will implement the model validation test
of Section~\ref{sec:validation}.\\
We assume that the vector $\mathbf{X}$ of gene expressions
defined by~\eqref{eq:def-X} follows a multivariate normal distribution
 $\mathcal{N}(m,\Sigma)$ where
$m=(m_{ik})_{1\leq i\leq N,1\leq k\leq K}$ is the mean and  $\Sigma=(\sigma_{ikj\ell})_{1\leq i,j\leq N,1\leq k,\ell\leq K}$ the covariance matrix that is supposed to be positive definite.
Under these conditions, $\mathbb{P}_\theta=\mathcal{N}(m,\Sigma)$ where the model parameter is $\theta=(m,\Sigma)$.  The precision
matrix is $\Sigma^{-1}$ and it is well-known (cf.\ e.g.~\citep[Eq.\,(27.6)]{Kendall1973}) that the
partial correlation matrix is given by % we define the matrix $C=(c_{ikj\ell})_{1\leq i,j\leq N,1\leq k,\ell\leq K}$ as
\[
C=- D\Sigma^{-1}D,\quad\text{where}\quad D=\text{diag}\left(\frac{1}{\sqrt{\Sigma^{-1}_{11}}},\ldots, \frac{1}{\sqrt{\Sigma^{-1}_{NN}}}\right).
\]
% It is known that $C$ is the partial correlation matrix (cf. annexe~\ref{sec:annexe}) defined as
% \[
%   c_{ikj\ell}=\text{cor}\left(X_{ik},X_{j\ell}\ \left|\ \left(X_{i'k'}\right)_{(i',k')\neq(i,k),(j,\ell)}\right.\right).
% \]
% This quantity is the correlation of the residuals in the regressions of $X_{ik}$ and $X_{j\ell}$ with respect to the other variables $(X_{i'k'})_{(i',k')\neq(i,k),(j,\ell)}$
% $(X_{i'k'})_{(i',k')\neq(i,k),(j,\ell)}$
% \textcolor{blue}{ligne trop longue} (cf. Proposition \ref{prop:precision-matrix}).
The gene network is the graph $G=(E,V)$ where the set $E$ of vertices
is made up of gene-time pairs $(i,k)$ where $1\leq i\leq N$ and $1\leq k\leq K$, and $V$  is the set of pairs $\{(i,k),(j,\ell)\}$ such that  $c_{ikj\ell}\neq 0$. In other words, gene $i$ at time $t_k$ is linked to gene $j$  at time
$t_\ell$ if and only $X_{ik}$ et $X_{j\ell}$   are not conditionally independent given all other gene expressions.
% A mechanistic interpretation of the edge is only possible when $k\not=\ell$.
% That is why it is in general not possible to define the model as mechanistic.\\

For simplicity, we consider the alteration of gene 1 of parameter $\beta\geq 0$. Since the model is
non-mechanistic, only the conditioning approach of Section~\ref{sec:cond} is relevant.
% Using the notations from Section~\ref{sec:general}, we denote the resulting law by $\mathbb{P}^\text{alt}_{(m,\Sigma)}$.
To simplify notations, we order pairs $(i,k)\in\{1,\ldots,N\}\times\{1,\ldots,K\}$ according to the lexicographic order and we
decompose by blocks the matrix $\Sigma$ and the vector $m$ as follows
\[
  \Sigma=
  \begin{pmatrix}
    \Sigma_{11} & \Sigma_{12} \\ \Sigma_{21} & \Sigma_{22}
  \end{pmatrix}\quad\text{and}\quad m=
  \begin{pmatrix}
    m_1 \\ m_2
  \end{pmatrix},
\]
where $\Sigma_{11}$ (resp.\ $\Sigma_{12}$, $\Sigma_{21}$, $\Sigma_{22}$) is a  matrix $K\times K$ (resp.\ $K\times (N-1)K$,
$(N-1)K\times K$, $(N-1)K\times (N-1)K$) and $m_1$ (resp.\ $m_2$) is a  $K$ (resp.\ $(N-1)K$)-dimensional vector.

\begin{prop}
  \label{prop:GGM}
  \begin{description}
  \item[\textmd{(i)}] The law $\mathbb{P}^{\text{alt},\beta}_{(m,\Sigma)}$ is Gaussian
    \[
      \mathbb{P}^{\text{alt},\beta}_{(m,\Sigma)}=\mathcal{N}\left(
        \begin{pmatrix}
          \beta m_1 \\ \mu(\beta)
        \end{pmatrix},
        \begin{pmatrix}
          \beta^2 \Sigma_{11} & \beta^2\Sigma_{12} \\ \beta^2\Sigma_{21} & \Gamma(\beta)
        \end{pmatrix}
      \right)
    \]
    where
    \begin{equation}
      \label{eq:def-mu}
      \mu(\beta)=m_2-(1-\beta)\Sigma_{21}\Sigma_{11}^{-1}m_1
    \end{equation}
    and
    \begin{equation}
      \label{eq:def-Gamma}
      \Gamma(\beta)=\Sigma_{22}-(1-\beta^2)\Sigma_{21}\Sigma_{11}^{-1}\Sigma_{12}.
    \end{equation}
  \item[\textmd{(ii)}] In addition, $\Gamma(0)^{-1}=(\Sigma^{-1})_{22}$.
  \end{description}
\end{prop}
\noindent Proposition~\ref{prop:GGM} will be proved in Appendix ~\ref{sec:annexe}.\\
Note that, if $\beta=0$, the \emph{alt} model is given by
\[
  \mathbb{P}^{\text{alt},0}_{(m,\Sigma)}=\mathcal{N}\left(
    \begin{pmatrix}
      0 \\ \mu(0)
    \end{pmatrix},
    \begin{pmatrix}
      0 & 0 \\ 0 & \Gamma(0).
    \end{pmatrix}
  \right)
\]
In particular, Item~(ii) means that the gene network of the knock-out model ($\beta=0$) is identical to the initial network in which
gene 1 and all incidental edges are deleted.

The previous result allows us to explicitly determine the sets of up-regulated, down-regulated and constant gene-date pairs defined
in~\eqref{eq:def-UDC} which take the form, after some algebra:
\begin{align*}
  \mathcal{U} & :=\left\{
    (i,k)\in\{2,\ldots,N\}\times\{2,\ldots,K\}: \varepsilon (\hat{m}_2)_{i,k}\leq -(\hat{\Sigma}_{21}\hat{\Sigma}_{11}^{-1}\hat{m}_1)_{i,k}
  \right\}, \\
  \mathcal{D} & :=\left\{
                (i,k)\in\{2,\ldots,N\}\times\{2,\ldots,K\}: \varepsilon (\hat{m}_2)_{i,k}\leq(\hat{\Sigma}_{21}\hat{\Sigma}_{11}^{-1}\hat{m}_1)_{i,k}
  \right\}, \\
  \mathcal{C} & :=\left\{
                (i,k)\in\{2,\ldots,N\}\times\{2,\ldots,K\}: \right. \\ & \qquad\qquad\qquad \qquad\qquad\qquad \left. -\varepsilon (\hat{m}_2)_{i,k}<(\hat{\Sigma}_{21}\hat{\Sigma}_{11}^{-1}\hat{m}_1)_{i,k}<\varepsilon (\hat{m}_2)_{i,k}
  \right\},
\end{align*}
where $\varepsilon>0$ and $\hat{\theta}_P=(\hat{m},\hat{\Sigma})$
% Notons en particulier que, lorsque $(m_2)_{i,k}>0$, $(i,k)\not\in\mathcal{U}$ si $(\Sigma_{21}\Sigma_{11}^{-1}m_1)_{i,k}\geq 0$, et
% $(i,k)\not\in\mathcal{D}$ sinon.
%
% \textcolor{blue}{je ne vois pas l'int\'er\^et ?}
%
% De plus, lorsque $(m_2)_{i,k}\leq 0$, $(i,k)\not\in\mathcal{C}$.
%
% \textcolor{blue}{pas clair pour moi }
%
% \textcolor{blue}{je propose de supprimer ce qui suit car on l'a d\'ej\`a dit\\
% $\hat{\theta}_P=(\hat{m},\hat{\Sigma})$ les param\`etres du mod\`ele graphique gaussien estim\'es sur le dataset initial. }
%
and the test statistic $S$ defined by~\eqref{eq:statistique-S} is constructed from
\begin{align*}
  \hat{\mathcal{U}}^\text{alt} & :=\left\{
    (i,k)\in\{2,\ldots,N\}\times\{2,\ldots,K\}: \bar{X}^\text{alt}_{i,k}\geq (1+\varepsilon)(\hat{m}_2)_{i,k}
  \right\}, \\
  \hat{\mathcal{D}}^\text{alt} & :=\left\{
    (i,k)\in\{2,\ldots,N\}\times\{2,\ldots,K\}: \bar{X}^\text{alt}_{i,k}\leq(1-\varepsilon) (\hat{m}_2)_{i,k}
  \right\}, \\
  \hat{\mathcal{C}}^\text{alt} & :=\Big\{
    (i,k)\in\{2,\ldots,N\}\times\{2,\ldots,K\}: \\ & \qquad\qquad\qquad\qquad
                                                     \left.(1-\varepsilon) (\hat{m}_2)_{i,k}<
                                                     \bar{X}^\text{alt}_{i,k}<(1+\varepsilon) (\hat{m}_2)_{i,k}
  \right\}.
\end{align*}
According to  Proposition~\ref{prop:GGM}, the law of the random vector $(\bar{X}^\text{alt}_{i,k})_{2\leq i\leq N,\,2\leq k\leq K}$ is
$\mathcal{N}\left(\hat{\mu},\frac{1}{P^\text{alt}}\hat{\Gamma}\right)$, where $\hat{\mu}$ and  $\hat{\Gamma}$  are obtained
from~\eqref{eq:def-mu} and~\eqref{eq:def-Gamma} with $(m,\Sigma)$ replaced by $(\hat{m},\hat{\Sigma})$. In particular, the $S$
statistic can be easily simulated to determine numerically the critical region.

Concerning the problem of improved parameter estimation from the initial dataset and an \emph{alt} dataset, as considered in
section~\ref{sec:affinage}, it seems difficult to construct a Gaussian graphical model able to account jointly for the two datasets,
because of the lack of mechanistic interpretation of the model.

\subsection{Mechanistic models}
\label{sec:GNI-mecaniste}

% Many mechanistic gene expression models were developed in the literature. The most common are Boolean
% networks,  those based on differential equations, Bayesian networks and penalized linear regression models. We do not discuss here
% Boolean networks because they were already studied from the perspective of \textcolor{blue}{biological intervention }in~\citep{Zhu2016,Azuaje2010,Ideker1999} and they do not account for quantitative gene expressions.
% Below
In this section, we apply our \emph{alt} modeling approach to Bayesian networks in Section~\ref{sec:bayes} and penalized linear
regression models in Section~\ref{sec:lasso}.
%to the modelling of biological intervention experiments experiments introduced in Section~\ref{sec:general}.

% \textcolor{blue}{ j'ai supprimé : nous consid\`ererons dans la suite les mod\`eles qui nous paraissent les plus adapt\'es}

\subsubsection{Dynamic Bayesian networks}
\label{sec:bayes}

First we describe the model, then we apply our \emph{alt} scheme.

\paragraph{Definition of dynamic Bayesian networks} ~

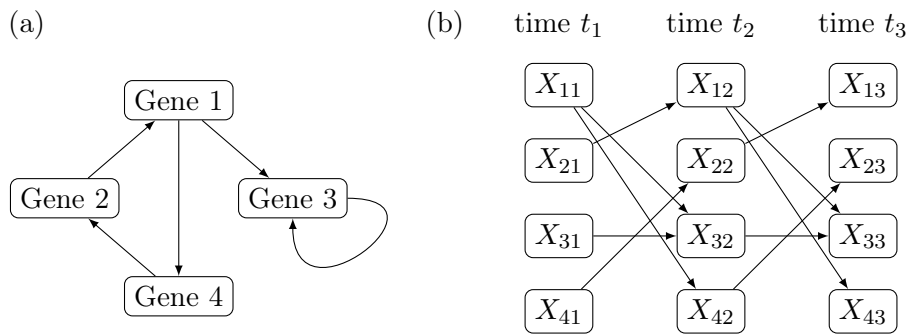
\begin{figure}[htp]
  \begin{center}
    \begin{tikzpicture}
      \node (a) at (-4,2.8) {(a)};
      \node[draw,rectangle,rounded corners=3pt] (G1) at (-2,1.8) {Gene 1};
      \node[draw,rectangle,rounded corners=3pt] (G2) at (-3.5,0.5) {Gene 2};
      \node[draw,rectangle,rounded corners=3pt] (G3) at (-0.5,0.5) {Gene 3};
      \node[draw,rectangle,rounded corners=3pt] (G4) at (-2,-0.8) {Gene 4};
      \draw[->,>=latex] (G2) -- (G1);
      \draw[->,>=latex] (G4) -- (G2);
      \draw[->,>=latex] (G1) -- (G3);
      \draw[->,>=latex] (G1) -- (G4);
      \draw[->,>=latex] (G3) to [out=0,in=270,looseness=5] (G3);
      \node (b) at (1.5,2.8) {(b)};
      \node (t1) at (3,2.8) {time $t_1$};
      \node[draw,rectangle,rounded corners=3pt] (X11) at (3,2) {$X_{11}$};
      \node[draw,rectangle,rounded corners=3pt] (X21) at (3,1) {$X_{21}$};
      \node[draw,rectangle,rounded corners=3pt] (X31) at (3,0) {$X_{31}$};
      \node[draw,rectangle,rounded corners=3pt] (X41) at (3,-1) {$X_{41}$};
      \node (t1) at (5,2.8) {time $t_2$};
      \node[draw,rectangle,rounded corners=3pt] (X12) at (5,2) {$X_{12}$};
      \node[draw,rectangle,rounded corners=3pt] (X22) at (5,1) {$X_{22}$};
      \node[draw,rectangle,rounded corners=3pt] (X32) at (5,0) {$X_{32}$};
      \node[draw,rectangle,rounded corners=3pt] (X42) at (5,-1) {$X_{42}$};
      \node (t1) at (7,2.8) {time $t_3$};
      \node[draw,rectangle,rounded corners=3pt] (X13) at (7,2) {$X_{13}$};
      \node[draw,rectangle,rounded corners=3pt] (X23) at (7,1) {$X_{23}$};
      \node[draw,rectangle,rounded corners=3pt] (X33) at (7,0) {$X_{33}$};
      \node[draw,rectangle,rounded corners=3pt] (X43) at (7,-1) {$X_{43}$};
      \draw[->,>=latex] (X21) -- (X12);
      \draw[->,>=latex] (X41) -- (X22);
      \draw[->,>=latex] (X11) -- (X32);
      \draw[->,>=latex] (X11) -- (X42);
      \draw[->,>=latex] (X31) -- (X32);
      \draw[->,>=latex] (X22) -- (X13);
      \draw[->,>=latex] (X42) -- (X23);
      \draw[->,>=latex] (X12) -- (X33);
      \draw[->,>=latex] (X12) -- (X43);
      \draw[->,>=latex] (X32) -- (X33);
    \end{tikzpicture}
  \end{center}
  \caption{Example of a dynamic Bayesian network with 4 genes and 3 times ($N=4$ and $K=3$): (a) static graph $G_\text{stat}$; (b)
    dynamic Bayesian dependency network $G$.}
  \label{fig:image7}
\end{figure}

In Bayesian networks, joint law of genes expression is expressed using an oriented acyclic graph $G$ representing the dependency
structure between genes~\citep{Hecker2009,Chang2011,Venkat2017}. In this section, we consider a particular class of dynamic Bayesian
networks, where the dependency structure between the variables $\mathbf{X}:=(X_{i,k})_{1\leq i\leq N,\,1\leq k<K}$ is encoded by a
static oriented graph $G_\text{stat}=(V_\text{stat},E_\text{stat})$ with $V_\text{stat}:=\{1,\ldots,N\}$ and $E_\text{stat}$ the set
of oriented edges of the graph, as shown in Figure~\ref{fig:image7}(a). From this static graph, the dynamic Bayesian network
$G=(V,E)$ with $V=\{1,\ldots,N\}\times\{1,\ldots,K\}$, where the set of edges $E$ is obtained as shown in Figure~\ref{fig:image7}(b),
so that a gene can only influence another gene (and possibly itself) at the time that immediately follows and that the parents of a
gene do not change over time. Note that the acyclicity condition is automatically satisfied.

% \textcolor{blue}{ Faut-il conserver : En
% particulier, des auto-régulations peuvent être incorporées au mod\`ele.}

Usually, the values taken by gene expressions are discretized in Bayesian networks. In general, this prevents to apply our method
with $\beta>0$. For $\beta=0$, we could consider any discretization of gene expressions containing value $0$. In the sequel, for
simplicity, we restrict to the case where gene expressions can take only two values 0 (for low expression) and 1 (for high
expression). In this case, the model corresponds to a probabilistic Boolean network. The probabilistic model corresponding to such a
dynamic Bayesian network is given by
\begin{align}
  \mathbb{P}(\mathbf{X}=\mathbf{x})=\prod_{k=1}^{K-1}\prod_{i=1}^{N}
  & \mathbb{P}\left(X_{i,k+1}=x_{i,k+1} \mid (X_{\ell,k})_{\ell \in A(i)}=(x_{\ell,k})_{\ell \in A(i)}\right) \notag \\
  & \times\prod_{i=1}^{N}\mathbb{P}(X_{i,1}=x_{i,1}),   \label{eq:reseau-bayes}
  % \label{eq3}
\end{align}
% \textcolor{blue}{formule \`a rediscuter}
where $\mathbf{x}=(x_{i,k})_{1\leq i\leq N,\, 1\leq k\leq K}\in\{0,1\}^{NK}$ and $A(i)$ is the set of parent nodes of $i\in \{1,\ldots,N\}$ in
the static graph.

The conditional probabilities of the dynamic Bayesian network are assumed to be independent of time. Thus, the parameters of the
model are
$\theta=(G,p^G)$ where $G$ is the static graph of the model and
\[
  p^G:=\left(p^G_{i}(x_i,(x_\ell)_{\ell\in A(i)}),\,1\leq i\leq N,\,x_i\in\{0,1\},\,(x_\ell)_{\ell\in A(i)}\in
    \{0,1\}^{\text{Card}(A(i))}\right),  % \label{eq:def-p_G} \\
\]
where, for all $i\in\{1,\ldots,N\}$,
\[
% p^G_{i}(x_i,(x_\ell)_{\ell\in A(i)})=\mathbb{P}\left(X_i=x_i \mid (X_\ell)_{\ell \in A(i)}=(x_\ell)_{\ell \in A(i)}\right)\in[0,1] \notag\
  p^G_{i}(x_i,(x_\ell)_{\ell\in A(i)}):=\mathbb{P}\left(X_{i,k+1}=x_i \mid (X_{\ell,k})_{\ell \in A(i)}=(x_\ell)_{\ell \in A(i)}\right)
\]
for all $k\in\{1,\ldots,K-1\}$ and
\[
  % 0\leq p^G_{i}(x_i,(x_\ell)_{\ell\in A(i)})\leq 1\quad\text{and}\quad
  p^G_{i}(0,(x_\ell)_{\ell\in A(i)})+p^G_{i}(1,(x_\ell)_{\ell\in A(i)})=1
\]
for all $(x_\ell)_{\ell\in A(i)}\in \{0,1\}^{\text{Card}(A(i))}$.

\paragraph{Description of conditional and mechanistic \emph{alt} models} ~

For simplicity, we consider the alteration of gene 1 of parameter $\beta = 0$, i.e.\ the knock-out of gene 1. From the model's
distribution~\eqref{eq:reseau-bayes}, it is straightforward to construct the alteration model using the conditioning approach of
Section~\ref{sec:cond}, denoted by $(X^\text{cond}_{ik})_{1\leq i\leq N,\,1\leq k\leq K}.$ The arguments developed to study the
example of Section~\ref{sec:exemple} can be easily extended to show that any dynamic Bayesian network is a mechanistic model. Hence
the mechanistic approach of Section~\ref{sec:meca} can also be applied to construct a gene alteration model
$(X^\text{meca}_{ik})_{1\leq i\leq N,\,1\leq k\leq K}$ as in Section~\ref{sec:exemple}.

% We denote by $(X^\text{cond}_{ik})_{1\leq i\leq N,\,1\leq k\leq K}$, resp.\
% $(X^\text{meca}_{ik})_{1\leq i\leq N,\,1\leq k\leq K}$ the gene expression values of the \textcolor{blue}{silencing model} obtained by the conditional
% (resp. mechanistic) method.
We denote by % $G_\text{stat}=(V_\text{stat},E_\text{stat})$ the static network associated with the dynamic
% Bayesian network $G=(V,E)$ (recall that $V=\{1,\ldots,N\}\times\{1,\ldots,K\}$ and $V_\text{stat}=\{1,\ldots,N\}$). We also denote by
$G'_\text{stat}$ the graph obtained from $G_\text{stat}$ by removing vertex 1 and all edges incident to 1. We call a
\emph{connected component} of $G_\text{stat}$ \emph{upstream} of gene 1 (resp. \emph{downstream} of gene 1) a connected component $\mathcal{C}$
of $G'_\text{stat}$ such that
\[
\forall j\in \mathcal{C},\quad (1,j)\not\in E_\text{stat} \quad(\text{resp.\ }(j,1)\not\in E_\text{stat}).
\]
According to the acyclicity assumption of the Bayesian network, a connected component ${\cal C}$ of $G'_\text{stat}$ is necessarily either upstream
or downstream of gene 1, or both when there is no edge connecting gene 1 to ${\cal C}$.

\begin{prop}
  \label{prop:bayes-1}
  Let $\mathcal{C}$  be a connected component downstream of gene 1. Then
  \begin{equation}
    \label{eq:thm-bayes-1}
    (X^\textnormal{cond}_{ik})_{i\in \mathcal{C},\,1\leq k\leq K} \stackrel{(d)}{=} (X^\textnormal{meca}_{ik})_{i\in \mathcal{C},\,1\leq k\leq K}
  \end{equation}
  and these two vectors have the same law as the random vector $ (X_{ik})_{i\in \mathcal{C},\,1\leq k\leq K}$ conditionally on $(X_{1k})_{1\leq k\leq K}=0$.\\
  In addition,
  % \textcolor{blue}{ Il me semble que, dans le membre de droite, c'est $X_{ik}$ au lieu de $X^\textnormal{meca}_{ik}$ ? }
  %  and
   $(X^\textnormal{cond}_{ik})_{i\in \mathcal{C},\,1\leq k\leq K}$ is independent of
  $(X^\textnormal{cond}_{ik})_{i\in \{2,\ldots,N\}\setminus\mathcal{C},\,1\leq k\leq K}$.
  An identical result holds true for  $X^\textnormal{meca}$.
\end{prop}

\begin{prop}
  \label{prop:bayes-2}
  Let $\mathcal{C}$  be a connected component upstream of gene 1. Then
  \begin{equation}
    \label{eq:bayes-2}
    (X^\textnormal{meca}_{ik})_{i\in \mathcal{C},\,1\leq k\leq K} \stackrel{(d)}{=} (X_{ik})_{i\in \mathcal{C},\,1\leq k\leq K}
  \end{equation}
   and $(X^\textnormal{meca}_{ik})_{i\in \mathcal{C},\,1\leq k\leq K}$ is independent of
  $(X^\textnormal{meca}_{ik})_{i\in \{2,\ldots,N\}\setminus\mathcal{C},\,1\leq k\leq K}$.
\end{prop}

\noindent
The proof of these results is given in Appendix~\ref{sec:bayes-annexe}. % These results assume \textcolor{blue}{total silencing of gene 1. We could extend
% them to the case of partial silencing} of gene 1, but the results would be weaker: \eqref{eq:thm-bayes-1}
% and~\eqref{eq:bayes-2} would remain true but the independence properties would fail. We leave the details to the interested reader.

We have already given in Section~\ref{sec:exemple} an example where the law of genes in a connected component upstream of gene 1 is
not the same in the conditional and the mechanistic approaches. In particular, the conditional model does not satisfy the
relationship~\eqref{eq:bayes-2} in general. Similarly, the independence property of Proposition \ref{prop:bayes-2} is not true in
general for the conditional model,, as the following example shows.

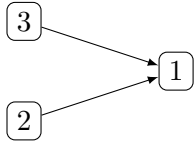
\begin{figure}[h]
  \begin{center}
    \begin{tikzpicture}
      \node[draw,rectangle,rounded corners=3pt] (X2) at (0,-0.66) {$2$};
      \node[draw,rectangle,rounded corners=3pt] (X3) at (0,0.66) {$3$};
      \node[draw,rectangle,rounded corners=3pt] (X1) at (2,0) {$1$};
      \draw[->,>=latex] (X2) -- (X1);
      \draw[->,>=latex] (X3) -- (X1);
    \end{tikzpicture}
  \end{center}
  \caption{Example of a static network where the conditional independence property is not satisfied by the conditional alt model.}
  \label{fig:exemple-2}
\end{figure}

We consider the gene expression model with $N=3$ and $K=2$ where $X_{11}$, $X_{21}$, $X_{22}$,
$X_{31}$ and $X_{32}$ are i.i.d. with Bernoulli distribution with parameter $1/2$ and
\[
  X_{12}=|X_{21}-X_{31}|.
\]
It is easy to verify that this model is a dynamic Bayesian network, the static network of which is shown in
Figure~\ref{fig:exemple-2}.

The sets $\{2\}$ are $\{3\}$ are two connected components upstream gene 1. Since
\[
\{X_{12}=0\}=\{X_{21}=X_{31}=1\}\cup\{X_{21}=X_{31}=0\}
\]
then conditionally on  $X_{12}=0$, the random variables $X_{21}$ are $X_{31}$ not independent (they are actually equal). This shows that the independance
property in Proposition \ref{prop:bayes-2} is not satisfied.

\paragraph{Improved estimation of initial model parameters using \emph{alt} data} ~

% The implementation of the validation test described in Section~\ref{sec:validation} does not pose any difficulty in the case of the Bayesian network models considered here.
% Details are left to the reader.\\
Bayesian networks are usually inferred by \emph{a posteriori} likelihood maximization on graphs. Let us recall the method. Let
$(\mathbb{Q}(G),\, G\in \mathcal{G})$ be an \emph{a priori} distribution over all oriented acyclic graphs and a dataset
$\mathbf{x}\in D^{NKP}$ as in~\eqref{eq:def-dataset-initial}. We maximise the \emph{a posteriori} log-likelihood with respect to
$(G,p^G)$:
\[
  \mathcal{L}(G,p^G\mid \mathbf{x}):=\log  \mathbb{P}_{(G,p^G)}(\mathbf{X}=\mathbf{x})+\log \mathbb{Q}(G).
\]
Classical choices for the \emph{a priori} distribution $\mathbb{Q}$ are those leading to an AIC or BIC type penalty in order to
select
networks with few edges \citep{Filkov2005}.\\
% This maximization is an NP-hard problem when all possible graphs are considered. It can be made easier, for example, by limiting the
% number of parents for each node, or by using heuristic \textit{hill-climbing} with random initialization \citep{Hartemink2000,Yu2002}.
% However, a large amount of data is required to make such an inference.\\
If an \emph{alt} dataset $\mathbf{x}^\text{alt}$ corresponding to a total silencing experiment of gene 1 is available, the method of
improvement of the parameter estimation of the initial model proposed in Section~\ref{sec:affinage} can be applied. Using the
previous notations and~\eqref{eq:amelioration-estimation}, the new \emph{a posteriori} log-likelihood to be maximized is expressed as
\[
  \mathcal{L}^\text{cond}(G,p^G\mid \mathbf{x},\mathbf{x}^\text{alt}):=\mathcal{L}(G,p^G\mid \mathbf{x})+
  % \log \mathbb{P}_{(G,p^G)}(\mathbf{X}=\mathbf{x})+
  \log\mathbb{P}^\text{alt}_{(G,p^G)}(\mathbf{X}^\text{cond}=\mathbf{x}^\text{alt})
  % +\log \mathbb{Q}(G).
\]
for the conditional model and
\[
  \mathcal{L}^\text{meca}(G,p^G\mid \mathbf{x},\mathbf{x}^\text{alt}):=\mathcal{L}(G,p^G\mid \mathbf{x})+
  % \log \mathbb{P}_{(G,p^G)}(\mathbf{X}=\mathbf{x})+
  \log\mathbb{P}^\text{alt}_{(G,p^G)}(\mathbf{X}^\text{meca}=\mathbf{x}^\text{alt}). % +\log \mathbb{Q}(G).
\]
for the mechanistic model.

\subsubsection{Penalized linear regression}
\label{sec:lasso}

We consider the linear, mechanistic, Gaussian model with delay (called LiRE model in~\citep{Rodolphe2025}), which generalizes the one
introduced in~\citep{Schleiss2021}. This model can be seen either as the discretization of a model based on differential equations or as
a regression model.

% \textcolor{blue}{faut-il indiquer :\\
% 1) qu'on peut voir soit comme la discr\'etisation d'un
% mod\`ele bas\'e sur des \'equations diff\'erentielles, soit comme un mod\`ele de r\'egression. \\
% 2)
% (Lin\'eaires \`a Retard Echelonn\'e), qui n'a plus de sens en anglais?}

\paragraph{Definition of the LiRE model} ~

In this model, gene expressions satisfy the following inductive relationships:
\begin{align}
X_{ik} &:= \mu_{ik} +
\sum_{j=1}^{N}  \sum_{\ell=1}^{k-1}
a_{\ell} \omega_{ij} X_{j (k-\ell)}+G_{ik}
\label{Eq:Probabilistic_model_one_time}
\end{align}
for all gene $i\in\{1,\ldots,N\}$ and time $k\in\{1,\ldots,K\}$, where $\mu_{ik}$ is the average expression of gene $i$ at time $k$,
in the absence of influence from other genes, $\omega_{ij}$ is the intrinsic (time-independent) force exerted by gene $j$
on the expression of gene $i$, $a_\ell$ is a (gene independent) coefficient that only depends on time differences and $G_{ik}$ stands for the noise.\\
The coefficients $a_\ell$ are assumed to be nonnegative, but $\omega_{ij}$ can be positive (resp.\ negative) if gene $j$ activates
(resp.\ inhibits) gene $i$ and null if gene $j$ has no influence on gene $i$. The term $a_\ell \omega_{ij}$ is the contribution made
by gene $j$ at time $k-\ell$ to the expression of gene $i$ at time $k$. This particular form dissociates the effects of time from
inter-gene effects. It is justified by the fact that the measurement dates $t_1,\ldots,t_K$ are regularly spaced. Replacing the
vector $(a_\ell)_\ell$ by a matrix of parameters depending both on the delay $\ell$ and the measurement time $k$ allows to consider
the case where the times of measurement are not regularly distributed~\citep{Schleiss2021}.

The associated gene network is the directed graph in which an edge points from $j$ to $i$ if and only $\omega_{ij}\neq 0$. Finally,
the random variables $(G_{ik})_{1\leq i\leq N,\,1\leq k\leq K}$ are assumed to be Gaussian centered independent, and the variance
$\sigma^2_i$ of $G_{ik}$ does not depend on $k$. In particular, the initial expressions
$(X_{i1})_{1\leq i\leq N}=(\mu_{i1}+G_{i1})_{1\leq i\leq N}$ of the genes are independent. We set:
\[
\Sigma:=
  \begin{pmatrix}
    \sigma_1^2 & \hdots & 0 \\
    \vdots & \ddots & \vdots \\
    0 & \hdots & \sigma_N^2
  \end{pmatrix}
\]
and
\[
\Lambda := \operatorname{diag}( \underbrace{
\Sigma, \cdots,\Sigma
}_{K \mbox{ times}}).
\]
In order for the model to be identifiable, we assume that $\sum_{\ell=1}^{K-1}a_\ell=1$.
The parameter space of the model is then
$\Theta= \mathbb{R}^{NK} \times {\cal S}_{K-2} \times \left( \mathbb{R}^{N^2} \setminus \{0\} \right) \times
\left(\mathbb{R}_{+}^{\star}\right)^{N}$, where ${\cal S}_d$ is the simplex of dimension $d$, and
\[
  \theta= \left( (\mu_{ik})_{1\leq i\leq N,\, 1\leq
      k\leq K}, (a_\ell)_{1\leq \ell\leq K-1}, (\omega_{ij})_{1\leq i,j\leq N}, (\sigma_{i})_{1\leq i\leq N} \right).
\]
is an element of $\Theta$.\\
Recall that $\mathbb{P}_\theta$ represents the distribution of $\mathbf{X}=(X_{ik})_{1\leq i\leq N,\,1\leq k\leq K}$
 when the parameter is $\theta\in\Theta$.
In the following, we adopt the vector representation
\[
  \mathbf{X}=(X_{1,1}, \ldots, X_{N,1}, \ldots, X_{1,K}, \ldots, X_{N,K}).
\]
We also define
\[
  \mu=(\mu_{1,1}, \ldots, \mu_{N,1}, \ldots, \mu_{1,K}, \ldots, \mu_{N,K})
\]
and $H$ the matrix defined by blocks as follows
\[
H:=
   \begin{pmatrix}
   I & 0 & \hdots & 0 & 0 \\
   -a_1 \Omega & I & \hdots & 0 & 0 \\
   \vdots & \vdots & \ddots & \vdots & \vdots \\
   -a_{T-2} \Omega & -a_{T-3} \Omega & \hdots & I  & 0\\
   -a_{T-1} \Omega & -a_{T-2} \Omega & \hdots  & -a_1 \Omega & I
   \end{pmatrix},
\]
where $I$ denotes the $N\times N$ identity matrix and $\Omega=(\omega_{ij})_{1\leq i,j\leq N}$. The following result
characterizes the probability measure $\mathbb{P}_\theta$.

\begin{thm} \label{Th:Vecteurs_gaussiens_E_X}
 For any $\theta\in\Theta$, $\mathbb{P}_{\theta}$ is  the multivariate Gaussian distribution
\begin{align}
  \mathbb{P}_{\theta}
  = \mathcal{N}(\Upsilon,
  \Gamma) \ , \label{Eq:X_gaussien}
\end{align}
where
\begin{equation}
  \label{eq:Upsilon-Gamma}
  \Upsilon
  :=H^{-1} \mu\quad\text{et}\quad
  \Gamma:=H^{-1} \Lambda
  \left(H^{-1}\right)^{\top}.
\end{equation}
\end{thm}

\paragraph{Mechanistic model of gene alteration} ~

% {\color{magenta} changer tous les $\beta$ en $1-\beta$}

Consider the alteration of gene 1 with parameter $\beta$. The LiRE model is mechanistic, so we can apply the method
defined in Section~\ref{sec:meca}. Let $(X^\text{meca}_{ik})_{1\leq i\leq N,\,1\leq k\leq K}$ be the model obtained:
\begin{align}
  X_{ik}^\text{meca}
  = \mu_{ik} + \beta \sum_{\ell=1}^{k-1}
  a_\ell \omega_{i1}
  X_{1,k-\ell}^\text{meca}
  +\sum_{j=2}^{N}  \sum_{\ell=1}^{k-1}
  a_\ell \omega_{ij}
  X_{j,k-\ell}^\text{meca}
  + G_{ik}  \label{Eq:x_meca_ARN_transcript}
\end{align}
for any $i \in \{1,\ldots,N\}$ and $k\in \{1,\ldots,K\}$.\\
We recognize a LiRE model where the matrix $\Omega$  has been replaced by $\Omega D_\beta$, where  $D_\beta$ is the $NK\times NK$ matrix defined by
\begin{align}
D_\beta = \operatorname{diag}(v)
% \underbrace{ v, \ldots ,v}_{K \mbox{ fois}}
  , \label{Eq:D_alpha_t}
\end{align}
and $v= (\beta,1,\ldots,1)\in\mathbb{R}^N$. Other parameters $\mu$, $a$ and $\sigma$ remain unchanged. Hence, we deduce from
\eqref{Eq:x_meca_ARN_transcript} that the law $\mathbb{P}^\text{alt,$\beta$,meca}_\theta$ of the mechanistic model of alteration of
gene 1 with parameter $\beta$ is
\begin{equation}
  \label{eq:modele-LiRE-meca}
  \mathbb{P}^\text{alt,$\beta$,meca}_\theta
  =\mathbb{P}_{( a,\Omega D_\beta, \mu,\sigma)}.
\end{equation}
Relation \eqref{eq:modele-LiRE-meca} implies that the gene network associated with the mechanistic alteration model is the same as in
the initial model, except when $\beta=0$ (gene knock-out). In this case, all edges starting from gene 1 are deleted.

\paragraph{Conditional model of gene alteration} ~

It is convenient to operate with a vector $W=(W^{(1)},W^{(2)})$ whose coordinates are those of $\mathbf{X}$ permuted:
\[
  W^{(1)}=(X_{11},\ldots,X_{1K})\quad\text{et}\quad W^{(2)}=(X_{21},\ldots,X_{2K},X_{31},\ldots,X_{NK}).
\]
Alteration concerns only the $W^{(1)}$ component.\\
 According to Theorem~\ref{Th:Vecteurs_gaussiens_E_X},
\[W \sim \mathcal{N}
\left(
\Upsilon_W,\Gamma_W
\right),\]
where
\[
\Upsilon_W=\left(\Upsilon^{(1)}_{W},\Upsilon^{(2)}_{W}\right) \quad\mbox{and}\quad
\Gamma_W=\left[\begin{array}{cc}
   \Gamma_W^{(11)} & \Gamma_W^{(12)} \\
   \Gamma_W^{(21)} & \Gamma_W^{(22)}
\end{array}\right]
\]
are obtained by permuting the rows and columns of $\Upsilon$ et $\Gamma$, $\Upsilon_W^{(1)}\in\mathbb{R}^K$,
$\Upsilon_W^{(2)}\in\mathbb{R}^{(N-1)K}$ and $\Gamma_W^{(11)}$ (resp.\  $\Gamma_W^{(12)}$, $\Gamma_W^{(21)}$ and $\Gamma_W^{(22)}$)
is a $K\times K$  (resp.\ $K\times
(N-1)K$, $(N-1)K\times K$ and $(N-1)K\times (N-1)K$) matrix.\\
Let us denote by
\[
{W}^{\text{cond}}
:= \left( {W}^{\text{cond},1},
{W}^{\text{cond},2} \right)
\]
The random vector obtained in the alteration of gene 1 with parameter $\beta$ using the conditional model of
Section~\ref{sec:cond}.%  Set
% \[
%   A_{\beta}:= (1-\beta)\Gamma^{(21)}_W \left(\Gamma^{(11)}_W\right)^{-1},
% \]
% where
% \[
%   \Gamma_W^{(2|1)}:= \Gamma^{(22)}_W - \Gamma^{(21)}_W \left(\Gamma^{(11)}_W\right)^{-1} \Gamma^{(12)}_W.
% \]

% \begin{rem}
%   \label{rem:Gamma_11-inversible}
%  The matrix $\Gamma^{(11)}_W$ is invertible because it is the covariance matrix of the Gaussian vector $W^{(1)}$.\\
%  For any
%   $k\in\{1,\ldots,K\}$, the random vector $X_{1k}$ can be written as $X_{1k}=G_{1k}+Y_{1k}$ where $Y_{1k}$ is independent of $G_{1k}$,   $G_{1k}\sim\mathcal{N}(0,\sigma_1^2)$  and
%   $\sigma_1>0$. We deduce, by recurrence on $k$ that $(X_{11},\ldots,X_{1k})$  has a density with respect to the Lebesgue measure.
% \end{rem}

\begin{thm}
  \label{Th:Silencing_conditionnal}
   Let $\mathbb{P}^{\text{alt},\beta,\text{cond}}_\theta$ denote the law of  $W^{\text{cond}}$, then
  \[
    \mathbb{P}^{\text{alt},\beta,\text{cond}}_{\theta}
    = \mathcal{N}\left(
      \Upsilon_{W}^{\text{cond}},
      \Gamma_{W}^{\text{cond}}
    \right)
  \]
  where
  \begin{equation}
    \label{eq:Upsilon-cond}
    \Upsilon_{W}^{\text{cond}}=\left(\beta \Upsilon^{(1)}_{W}\, ,\, \Upsilon_{W}^{(2)} -(1-\beta) \Gamma^{(21)}_W
      \left(\Gamma^{(11)}_W\right)^{-1} \Upsilon^{(1)}_{W}\right)
  \end{equation}
  and
  \begin{equation}
    \label{eq:Gamma-cond}
    \Gamma_{W}^{\text{cond}}=
    \begin{pmatrix}
      \beta^2 \Gamma^{(11)}_W & \beta^2 \Gamma^{(12)}_W \\
      \beta^2 \Gamma^{(21)}_W & \Gamma_W^{(22)}-(1-\beta^2)\Gamma^{(21)}_W\left(\Gamma^{(11)}_W\right)^{-1}\Gamma^{(12)}_W
    \end{pmatrix}.
  \end{equation}
\end{thm}
We recover a similar expression as in Proposition~\ref{prop:GGM}.
This result is proven in the more general context of simultaneous alteration of several genes in~\citep{Rodolphe2025}.
% It could be that the conditional model is a LiRE model. This would require checking that ~\eqref{eq:Upsilon-cond} and~\eqref{eq:Gamma-cond} can be written as
% in the form~\eqref{eq:Upsilon-Gamma}.
Note that the conditional \emph{alt} model of the last theorem does not seem to be a LiRE model. It is therefore not obvious to associate a
gene network with the conditional \emph{alt} model that can be compared with that of the initial LiRE model.

\paragraph{Improved estimation of initial model parameters using \emph{alt} data} ~

The method proposed in~\citep{Rodolphe2025} for
estimating the parameters of the LiRE model is based on a posteriori likelihood maximization with the prior
density:

\begin{align}
\pi(\theta) =
\prod_{i=1}^{N} \left(\left(\frac{\tau_i}{2\sigma_i^2}\right)^N
\exp \left(-
\frac{\tau_i}{\sigma_i^2}\sum_{j=1}^N|\omega_{ij}|
\right)\right),
\end{align}
where $\tau=(\tau_1,\cdots,\tau_N) \in \left(\mathbb{R}^{\star}_{+}\right)^{N}$ is a parameter defined by the user.
In other words, conditionally on $\sigma$, the $\omega_{ij}$ follow independent Laplace laws. A flat a priori is assumed for the other parameters $\mu$, $a$ and $\sigma$.\\
Therefore, denoting by $f_\theta$ the density of the distribution of $\mathbf{X}$ under $\mathbb{P}_\theta$, inference of the parameter $\theta$
is performed by maximizing the a posteriori log-likelihood of observations $\mathbf{x}$.
\begin{align}
  & \mathcal{L}(\theta\mid \mathbf{x}) =\sum_{p=1}^{P}\log f_\theta\big((x_{i,k,p})_{1\leq i\leq N,\,1\leq k\leq
    K}\big)+\log\pi(\theta) \label{eq:vrais-LiRE} \\
                                     &= - \frac{1}{2}
\sum_{i=1}^{N} \frac{1}{\sigma_i^{2}} \left(
\sum_{k=1}^K\sum_{p=1}^{P}\left(x_{i,k,p} - \mu_{ik}-
\sum_{j=1}^{N}\sum_{\ell=1}^{k-1}a_\ell \omega_{ij} x_{j,k-\ell,p}\right)^{2}
+ 2\tau_i \sum_{j=1}^N|\omega_{ij}|
\right) \notag \\
&  \phantom{=}- \left(\frac{PK+2N}{2}\right) \sum_{i=1}^{N} \ln \sigma_i^2
+ N\sum_{i=1}^{N} \ln\frac{\tau_i}{2}. \notag
\end{align}
The maximization algorithm of $\mathcal{L}(\theta\mid\mathbf{x})$ used in~\citep{Rodolphe2025} consists in successively optimizing
on the 4 parameters $\mu$, $\sigma$, $a$ and $\Omega$ while fixing the others. Optimization in $\mu$ and $\sigma$ is explicit, that
in $a$ can be expressed as a least-squares problem with a positivity constraint and that in $\Omega$ can be reduced to $N$ lasso
problems~\citep{Tibshirani1996}, each involving a vector of the form $(\omega_{i1},\ldots,\omega_{iN})$.
The choice of the parameters $\tau_i$ can be made by cross-validation.

% Since the laws of the two alt models above are Gaussian and explicit, the validation test described in Section~\ref{sec:validation} can
% be carried out in the same way as for Gaussian graphical models (see Section~\ref{sec:GGM}). In the sequel, we focus on the problem of
% improvement of estimation from a alt dataset:
If we are given an additional dataset $\mathbf{x}^\text{alt}$ of size $P^\text{alt}$ from an alteration experiment of gene 1 with
parameter $\beta$, we can apply the method in Section~\ref{sec:affinage}. Let $f_\theta^{\text{alt,}\beta\text{,meca}}$ stand for the
density of $\mathbf{X}$ under $\mathbb{P}^\text{alt,$\beta$,meca}_\theta$. Using~\eqref{eq:vrais-LiRE}, the new a posteriori
log-likelihood is expressed in the case of the mechanistic approach as
\begin{multline*}
  \mathcal{L}^\text{meca}(\theta\mid \mathbf{x},\mathbf{x}^\text{alt})=\sum_{p=1}^P\log f_\theta\big((x_{i,k,p})_{1\leq i\leq N,\,1\leq k\leq K}\big)
  \\ +\sum_{p=1}^{P^\text{alt}}\log f_\theta^{\text{alt,}\beta\text{,meca}}\left((x^\text{alt}_{i,k,p})_{1\leq i\leq N,\,1\leq k\leq K}\right)+\log\pi(\theta).
\end{multline*}
Using~\eqref{eq:modele-LiRE-meca} and Theorem~\ref{Th:Vecteurs_gaussiens_E_X}, we obtain
\begin{align*}
  & \mathcal{L}^\text{meca}(\theta\mid \mathbf{x},\mathbf{x}^\text{alt}) = - \frac{1}{2}
\sum_{i=1}^{N} \frac{1}{\sigma_i^{2}} \left(
\sum_{k=1}^K\sum_{p=1}^{P}\left(x_{i,k,p} - \mu_{ik}-
    \sum_{j=1}^{N}\sum_{\ell=1}^{k-1}a_\ell \omega_{ij} x_{j,k-\ell,p}\right)^{2} \right. \\
  & +\sum_{k=1}^K\sum_{p=1}^{P^\text{alt}}\left(x^\text{alt}_{i,k,p} - \mu_{ik}-\beta\sum_{\ell=1}^{k-1}a_\ell \omega_{i1} x^\text{alt}_{1,k-\ell,p}-
    \sum_{j=2}^{N}\sum_{\ell=1}^{k-1}a_\ell \omega_{ij} x^\text{alt}_{j,k-\ell,p}\right)^{2} \\
  & \left. + 2\tau_i \sum_{j=1}^N|\omega_{ij}|\right)- \left(\frac{PK+P^\text{alt}K+2N}{2}\right) \sum_{i=1}^{N} \ln \sigma_i^2
    + N\sum_{i=1}^{N} \ln\frac{\tau_i}{2}.
\end{align*}
% \textcolor{blue}{Je pense que la formule n'est pas exacte, \`a rediscuter}

Maximization of this log-likelihood can be achieved using the same iterative method as for inference: explicit optimization in $\mu$ and $\sigma$, optimization in $a$ by least squares with a positivity
constraint and optimization in $\Omega$ by \emph{lasso}.

In the case of the conditional approach, % let $f_\theta^{\text{BI,}\beta\text{,cond}}$  be the density of $\mathbf{X}$ under
% $\mathbb{Q}^\text{BI,$\beta$,cond}_\theta$.
% The aim is to maximize
the log-likelihood
% \begin{multline*}
%   \mathcal{L}^\text{cond}(\theta\mid \mathbf{x},\mathbf{x}^\text{BI})=\sum_{p=1}^P\log f_\theta\big((x_{i,k,p})_{1\leq i\leq N,\,1\leq k\leq K}\big)
%   \\ +\sum_{p=1}^{P^\text{BI}}\log f_\theta^{\text{BI,}\beta\text{,cond}}\left((x^\text{BI}_{i,k,p})_{1\leq i\leq N,\,1\leq k\leq K}\right)+\log\pi(\theta).
% \end{multline*}
% \textcolor{blue}{ Il faut remplacer $G$ par $\theta$ dans le membre de droite ?}
can be computed explicitely according to Theorem~\ref{Th:Silencing_conditionnal}, but its expression lacks the structure that allows successive
optimization in each parameter as above. Hence likelihood maximization is difficult for this model.

\appendix

\section{Conditional laws of Gaussian vectors and proof of Proposition~\ref{prop:GGM}}
\label{sec:annexe}
\label{sec:preuve-GGM}

Let $m$ and $d$ denote integers such that $1\leq m< d$. We consider a Gaussian vector $Y\in\mathbb{R}^d$ with mean $\mu$ and
covariance matrix $\Gamma$. Write $Y=(Y_1,Y_2)$ where $Y_1\in\mathbb{R}^m$ is the vector of the first $m$ coordinates of $Y$ and
$Y_2\in\mathbb{R}^{d-m}$ is the vector of the last $d-m$ coordinates. In a similar way, we decompose the vector $\mu$ and the matrix
$\Gamma$:
\begin{equation}
  \label{eq:decomp-Gamma}
  \mu=(\mu_{1},\mu_{2}) \quad \text{et}\quad
  \Gamma=
  \begin{pmatrix}
    \Gamma_{11} & \Gamma_{12} \\ \Gamma_{21} & \Gamma_{22},
  \end{pmatrix}
\end{equation}
where $\mu_1\in\mathbb{R}^m$, $\mu_2\in\mathbb{R}^{d-m}$, $\Gamma_{11}$ (resp.\ $\Gamma_{12}$, $\Gamma_{21}$, $\Gamma_{22}$) is a
$m\times m$ (resp.\ $m\times(d-m)$, $(d-m)\times m$, $(d-m)\times(d-m)$) matrix. The following result describes the conditional
distribution of $Y_1$ given $Y_2$~\citep[Prop.\,3.13]{Eaton1983}.

\begin{thm}
  \label{thm:conditional_Gaussian} If $\Gamma$ is invertible, then for any $y_2\in\mathbb{R}^{d-m}$,
  \[
    \textnormal{Law} (Y_1 \mid Y_2=y_2)
    =\mathcal{N}_m(\mu_{1|2}(y_2) , \Gamma_{1|2} ),
  \]
  where
  \begin{align*}
    \mu_{1|2}(y_2) & = \mu_{1} + \Gamma_{12} \Gamma_{22}^{-1}(y_2-\mu_2) \\
    \Gamma_{1|2} & =\Gamma_{11}-\Gamma_{12} \Gamma_{22}^{-1}\Gamma_{21}.
  \end{align*}
\end{thm}
The case $\beta=0$ in Point~(i) of Proposition~\ref{prop:GGM} is an immediate consequence of the last formula
(exchanging the role of $Y_1$ and $Y_2$). We leave to the reader the general case.

The proof of Theorem~\ref{thm:conditional_Gaussian} is based on the calculation of the precision matrix $\Gamma^{-1}$ given by the
following proposition.

\begin{prop}
  \label{prop:precision-matrix} If $\Gamma$ is invertible, then
  \[
    \Gamma^{-1} =
    \begin{pmatrix}
      \Gamma_{1|2}^{-1} & -\Gamma_{1|2}^{-1} \Gamma_{12}\Gamma_{22}^{-1} \\
      - \Gamma_{22}^{-1} \Gamma_{21} \Gamma_{1|2}^{-1} & \Gamma_{22}^{-1}+
      \Gamma_{22}^{-1} \Gamma_{21}\Gamma_{1|2}^{-1} \Gamma_{12} \Gamma_{22}^{-1}
    \end{pmatrix}
  \]
  where $\Gamma_{2|1} =\Gamma_{22}-\Gamma_{21} \Gamma_{11}^{-1}\Gamma_{12}$.
\end{prop}

Writing as in~\eqref{eq:decomp-Gamma}
\[
  \Gamma^{-1}=
\begin{pmatrix}
  (\Gamma^{-1})_{11} & (\Gamma^{-1})_{12} \\ (\Gamma^{-1})_{21} & (\Gamma^{-1})_{22}
\end{pmatrix},
\]
we deduce from Proposition \ref{prop:precision-matrix} that $(\Gamma^{-1})_{11}=\Gamma_{1|2}^{-1}$, hence Point~(ii)
of Proposition~\ref{prop:GGM} follows.

\section{Proof Propositions~\ref{prop:bayes-1} and ~\ref{prop:bayes-2} }
\label{sec:bayes-annexe}

We consider only the case of a dynamic Bayesian model with discrete values (as usual for gene networks). A similar proof can be done
in the case where gene expressions are continuous variables.

\subsection{Proof of Proposition~\ref{prop:bayes-1}}
\label{sec:proof-prop-bayes-1}

Throughout, $\mathcal{C}$ denotes a connected component of $G_\text{stat}$ downstream of gene 1.
\medskip

\noindent\textit{Step 1. Study of the conditional model}

As defined by the Bayesian model
\begin{align*}
  \mathbb{P} & \left((X_{ik})_{i\in\mathcal{C},\, 1\leq k\leq K} =(x_{ik})_{i\in\mathcal{C},\, 1\leq k\leq K} ,\ (X_{1k})_{1\leq k\leq
      K}=(x_{1k})_{1\leq k\leq K},\right. \\ & \left. \qquad\qquad\qquad\qquad\qquad(X_{jk})_{j\not\in\mathcal{C}\cup\{1\},\, 1\leq k\leq
      K}=(x_{jk})_{j\not\in\mathcal{C}\cup\{1\},\, 1\leq k\leq K}\right) \\
 & =\mathbb{P}\left((X_{ik})_{i\in\mathcal{C},\, 1\leq k\leq K} =(x_{ik})_{i\in\mathcal{C},\, 1\leq k\leq K} \mid (X_{1k})_{1\leq k\leq
      K}=(x_{1k})_{1\leq k\leq K}\right) \\ & \times \mathbb{P}\left((X_{1k})_{1\leq k\leq
      K}=(x_{1k})_{1\leq k\leq K},\ (X_{jk})_{j\not\in\mathcal{C}\cup\{1\}),\, 1\leq k\leq
      K}=(x_{jk})_{j\not\in\mathcal{C}\cup\{1\},\, 1\leq k\leq K}\right)
\end{align*}
for all $(x_{1k})_{1\leq k\leq K}$, $(x_{ik})_{i\in\mathcal{C},\,
  1\leq k\leq K}$ and $(x_{jk})_{j\not\in\mathcal{C}\cup\{1\},\,
  1\leq k\leq K}$.
  Recall that the law of the conditional model $(X^\text{cond}_{ik})_{1\leq i\leq N,\,1\leq k\leq K}$ is the conditional law of $(X_{ik})_{2\leq i\leq N,\,1\leq k\leq K}$ given $(X_{1k})_{1\leq k\leq K}=0$.
  From the previous formula, we deduce that
  \begin{multline*}
  \mathbb{P} \left((X^\text{cond}_{ik})_{2\leq i\leq N,\, 1\leq k\leq K} =(x_{ik})_{2\leq i\leq N,\, 1\leq k\leq K}\right) \\
  \begin{aligned}
    & =\mathbb{P}\left((X_{ik})_{i\in\mathcal{C},\, 1\leq k\leq K} =(x_{ik})_{i\in\mathcal{C},\, 1\leq k\leq K} \mid (X_{1k})_{1\leq k\leq
      K}=0\right) \\ & \times \mathbb{P}\left((X_{jk})_{j\not\in\mathcal{C}\cup\{1\}),\, 1\leq k\leq
      K}=(x_{jk})_{j\not\in\mathcal{C}\cup\{1\},\, 1\leq k\leq K}\mid (X_{1k})_{1\leq k\leq
      K}=0\right).
  \end{aligned}
  \end{multline*}
Hence $(X^\text{cond}_{ik})_{i\in\mathcal{C},\, 1\leq k\leq K}$ is independent of  $(X^\text{cond}_{jk})_{j\not\in\mathcal{C}\cup\{1\},\, 1\leq k\leq
  K}$ and
\begin{multline*}
  \mathbb{P}\left((X^\text{cond}_{ik})_{i\in\mathcal{C},\, 1\leq k\leq K} =(x_{ik})_{i\in\mathcal{C},\, 1\leq k\leq K}\right) \\
  =\mathbb{P}\left((X_{ik})_{i\in\mathcal{C},\, 1\leq k\leq K} =(x_{ik})_{i\in\mathcal{C},\, 1\leq k\leq K} \mid (X_{1k})_{1\leq k\leq
      K}=0\right).
\end{multline*}
\medskip
\noindent\textit{Step 2. Study of the mechanistic model}

It is easy to show, as we did in Section~\ref{sec:exemple}, that the dynamic Bayesian model
model described in Section~\ref{sec:bayes} can be written as a mechanistic model of the form
\[
  X_{i,k+1}=F_{i,k}\big((X_{\ell,k})_{\ell\in A(i)};\,U_{i,k}\big),
\]
for all $i\in\{1,\ldots, N\}$ and $k\in\{1,\ldots, K-1\}$, where $F_{i,k}((x_{\ell,k})_{\ell\in A(i)};\cdot)$
is the generalized inverse distribution function
of the conditional law of $X_{i,k+1}$ given $(X_{\ell,k})_{\ell\in A(i)}=(x_{\ell,k})_{\ell\in A(i)}$ and
 the variables $U_{i,k}$ are i.i.d. uniform on $[0,1]$. We deduce that the mechanistic model for total silencing of gene 1 is given by
\begin{align}
  X^\text{meca}_{1,k+1} & =0, \notag \\
  X^\text{meca}_{i,1} & =X_{i1}, \label{eq:CI-bayes} \\
  X^\text{meca}_{i,k+1} & =\hat{F}_{i,k}\big((X^\text{meca}_{\ell,k})_{\ell\in A(i)\setminus\{1\}};\,U_{i,k}\big), \label{eq:rec-meca-sil}
\end{align}
for all  $i\in\{2,\ldots, N\}$ and $k\in\{1,\ldots, K-1\}$, where
\begin{equation}
  \label{eq:def-hat-F}
  \hat{F}_{i,k}\big((x_{\ell,k})_{\ell\in A(i)\setminus\{1\}};u_{ik}\big)=
  \begin{cases}
    F_{i,k}\big(0,(x_{\ell,k})_{\ell\in A(i)\setminus\{1\}};u_{ik}\big)
    & \text{if } 1\in A(i)\\
    F_{i,k}\big((x_{\ell,k})_{\ell\in A(i)};u_{ik}\big) & \text{if } 1\not\in A(i).
  \end{cases}
\end{equation}
%
% \bigskip
%
% \textcolor{blue}{je supprime la phrase:\\
% Let us begin by proving Proposition~\ref{prop:bayes-1}\\
% car nous avons changé la structure de la preuve.}
%
% \textcolor{blue}{je supprime :\\
%  Let $\mathcal{C}$ be a connected component of $G_\text{stat}$ downstream  gene 1.\\
%   car cette information a été donnée au début de la preuve.}
Since $A(i)\cap\mathcal{C}=\emptyset$ for all  $i\not\in\mathcal{C}\cup\{1\}$, the relations
\begin{equation}
  \label{eq:preuve-bayes}
  X^\text{meca}_{i,k+1} =\hat{F}_{i,k}\big((X^\text{meca}_{\ell,k})_{\ell\in A(i)\setminus\{1\}};\,U_{i,k}\big)
\end{equation}
satisfied for any $i\not\in\mathcal{C}\cup\{1\}$ and $1\leq k\leq K-1$
imply
\begin{equation}
  \label{eq:indep-1}
\sigma\left((X^\text{meca}_{ik})_{i\not\in\mathcal{C}\cup\{1\},\,1\leq k\leq K}\right)\subset
\sigma\left((X_{i1})_{i\not\in\mathcal{C}\cup\{1\}},
(U_{ik})_{i\not\in\mathcal{C}\cup\{1\},\,1\leq k\leq
  K-1}
\right)
\end{equation}
where $\sigma\big((Z_j)_{j\in J}\big)$ stands for the $\sigma$-algebra generated of the random variables $(Z_j)_{j\in J}$.

 Similarly, since for all $i\in\mathcal{C}$ we have  $A(i)\subset \mathcal{C}\cup\{1\}$ the relations~\eqref{eq:preuve-bayes} with
 $i\in\mathcal{C}$, $ 1\leq k\leq K-1$ imply
\begin{equation}
  \label{eq:indep-2}
\sigma\left((X^\text{meca}_{ik})_{i\in\mathcal{C},\,1\leq k\leq K}\right)\subset
\sigma\left((X_{i1})_{i\in\mathcal{C}},
(U_{ik})_{i\in\mathcal{C},\,1\leq k\leq
  K-1}
\right).
\end{equation}
Since the families of random variables $(U_{ik})_{1\leq i\leq N, 1\leq k\leq K}$, $(X_{i1})_{i\in\mathcal{C}}$ and

\noindent
$(X_{i1})_{i\not\in\mathcal{C}\cup\{1\}}$ are independent, we deduce that $(X^\text{meca}_{ik})_{i\in\mathcal{C},\, 1\leq k\leq K}$ is independent of
$(X^\text{meca}_{jk})_{j\not\in\mathcal{C}\cup\{1\},\, 1\leq k\leq K}$.
\medskip

\noindent\textit{Step 3. Conclusion}
To complete the proof of Proposition~\ref{prop:bayes-1}
it remains to show that
\begin{multline*}
  \mathbb{P}\left((X^\text{meca}_{ik})_{i\in\mathcal{C},\, 1\leq k\leq K} =(x_{ik})_{i\in\mathcal{C},\, 1\leq k\leq K}\right) \\
  =\mathbb{P}\left((X_{ik})_{i\in\mathcal{C},\, 1\leq k\leq K} =(x_{ik})_{i\in\mathcal{C},\, 1\leq k\leq K} \mid (X_{1k})_{1\leq k\leq
      K}=0\right).
\end{multline*}
We will establish this identity by showing by induction on $\ell$ belonging to $\{1,\ldots,K\}$
\begin{multline}
  \mathbb{P}\left((X^\text{meca}_{ik})_{i\in\mathcal{C},\, 1\leq k\leq \ell} =(x_{ik})_{i\in\mathcal{C},\, 1\leq k\leq \ell}\right) \\
  =\mathbb{P}\left((X_{ik})_{i\in\mathcal{C},\, 1\leq k\leq \ell} =(x_{ik})_{i\in\mathcal{C},\, 1\leq k\leq \ell} \mid (X_{1k})_{1\leq k\leq
      \ell}=0\right). \label{eq:formule-Pierre}
\end{multline}

The property is easily verified when
$\ell=1$ , using~\eqref{eq:CI-bayes} and the fact that the random variables $(X_{i1})_{1\leq i\leq N}$ are independent.\\
 Assume that~\eqref{eq:formule-Pierre} is true for
  $\ell\in\{1,\ldots, K-1\}$  and let us prove it for $\ell+1$.
 Using~\eqref{eq:rec-meca-sil},~\eqref{eq:def-hat-F} and~\eqref{eq:formule-Pierre}, we have for all  $(x_{ik})_{i\in\mathcal{C},\,
  1\leq k\leq \ell+1}$,
\begin{align*}
  \mathbb{P}
  & \left((X^\text{meca}_{ik})_{i\in\mathcal{C},\, 1\leq k\leq \ell+1} =(x_{ik})_{i\in\mathcal{C},\, 1\leq k\leq
    \ell+1}\right) \\
  & =\mathbb{P}\left((X^\text{meca}_{i,\ell+1})_{i\in\mathcal{C}} =(x_{i,\ell+1})_{i\in\mathcal{C}}\mid
    (X^\text{meca}_{ik})_{i\in\mathcal{C},\, 1\leq k\leq \ell} =(x_{ik})_{i\in\mathcal{C},\, 1\leq k\leq \ell}\right) \\
  & \qquad\qquad {\times} \mathbb{P}\left((X^\text{meca}_{ik})_{i\in\mathcal{C},\, 1\leq k\leq \ell} =(x_{ik})_{i\in\mathcal{C},\, 1\leq k\leq
    \ell}\right) \\
  & =\mathbb{P}\left({\hat{F}_{i,\ell}\big((X^\text{meca}_{j,\ell})_{j\in
    A(i)\setminus\{1\}};}\,U_{i\ell}\big)=x_{i,\ell+1}, {\forall}\,i\in\mathcal{C}\mid
    (X^\text{meca}_{i\ell})_{i\in\mathcal{C}} =(x_{i\ell})_{i\in\mathcal{C}}\right) \\
  & \qquad\qquad {\times} \mathbb{P}\left((X_{ik})_{i\in\mathcal{C},\, 1\leq k\leq \ell} =(x_{ik})_{i\in\mathcal{C},\, 1\leq k\leq \ell} \mid (X_{1k})_{1\leq k\leq
      \ell}=0\right) \\
  & =\mathbb{P}\left(\hat{F}_{i,\ell}\big((x_{j,\ell})_{j\in
    A(i)\setminus\{1\}};\,U_{i\ell}\big)=x_{i,\ell+1}, {\forall}\,i\in\mathcal{C}\right) \\
  & \qquad\qquad {\times} \mathbb{P}\left((X_{ik})_{i\in\mathcal{C},\, 1\leq k\leq \ell} =(x_{ik})_{i\in\mathcal{C},\, 1\leq k\leq \ell} \mid (X_{1k})_{1\leq k\leq
      \ell}=0\right) \\
  % & =\mathbb{P}\left(F_{i,\ell}\big((x_{j,\ell})_{\ell\in
  %   A(i)};\,U_{i\ell}\big)=x_{i,\ell+1},{\forall}\,i\in\mathcal{C}\right) \\
  % & \qquad\qquad {\times} \mathbb{P}\left((X_{ik})_{i\in\mathcal{C},\, 1\leq k\leq \ell} =(x_{ik})_{i\in\mathcal{C},\, 1\leq k\leq \ell} \mid (X_{1k})_{1\leq k\leq
  %   \ell}=0\right) \\
  & =\mathbb{P}\left( X_{i,\ell+1}=x_{i,\ell+1},{\forall}\,i\in\mathcal{C}\mid
    X_{1\ell}=0,\,(X_{i\ell})_{i\in\mathcal{C}} =(x_{i\ell})_{i\in\mathcal{C}}\right) \\
  & \qquad\qquad {\times} \mathbb{P}\left((X_{ik})_{i\in\mathcal{C},\, 1\leq k\leq \ell} =(x_{ik})_{i\in\mathcal{C},\, 1\leq k\leq \ell} \mid (X_{1k})_{1\leq k\leq
    \ell}=0\right) \\
  & =\mathbb{P}\left((X_{ik})_{i\in\mathcal{C},\, 1\leq k\leq \ell+1} =(x_{ik})_{i\in\mathcal{C},\, 1\leq k\leq \ell+1} \mid (X_{1k})_{1\leq k\leq
      \ell+1}=0\right),
\end{align*}
% \textcolor{blue}{ je ne comprends pas la derni\`ere \'egalit\'e, pourquoi $F_{i,\ell}$ disparait ?}
where the fourth equality follows from the definition of $\hat{F}_{i,\ell}$. %\textcolor{cyan}{[est-ce que c'est plus clair maintenant ?]}
This concludes the proof of Proposition~\ref{prop:bayes-1}.

\subsection{Proof of Proposition~\ref{prop:bayes-2}}
\label{sec:proof-prop-bayes-2}

Let $\mathcal{C}$  be a connected component of $G_\text{stat}$, upstream of gene 1. \\
Similarly as in Step 2 of the proof of Section~\ref{sec:proof-prop-bayes-1}, the $\sigma$-fields inclusions~\eqref{eq:indep-1}
and~\eqref{eq:indep-2} still hold true in this case. Hence $(X^\textnormal{meca}_{ik})_{i\in \mathcal{C},\,1\leq k\leq K}$ is independent of
$(X^\textnormal{meca}_{ik})_{i\in \{2,\ldots,N\}\setminus\mathcal{C},\,1\leq k\leq K}$. Furthermore, since, for all $i\in{\cal C}$
and $1\leq k\leq K$,
\[
\hat{F}_{i,k}\big((x_{\ell,k})_{\ell\in A(i)\setminus\{1\}};u_{ik}\big)=F_{i,k}\big((x_{\ell,k})_{\ell\in A(i)};u_{ik}\big),
\]
we deduce
\begin{multline*}
\mathbb{P}\left((X^\text{meca}_{ik})_{i\in\mathcal{C},\, 1\leq k\leq K} =(x_{ik})_{i\in\mathcal{C},\, 1\leq k\leq K}\right) \\
  =\mathbb{P}\left((X_{ik})_{i\in\mathcal{C},\, 1\leq k\leq K} =(x_{ik})_{i\in\mathcal{C},\, 1\leq k\leq K}\right).
\end{multline*}
Hence Proposition~\ref{prop:bayes-2} is proved.

\bigskip

\noindent\textbf{Acknowledgements:} We would like to thank Alvyn Bonnet who, as a Master student, carried out a major bibliographical
compilation. RL thanks Canc\'erop\^ole Est and R\'egion Grand Est for the funding of his PhD thesis. NC, LV and PV thank INSERM /
AVIESAN ITMO Cancer for the funding of the project Predi-CLL: ``Quantification et pr\'ediction de l'\'evolution de
l'h\'et\'erog\'en\'eit\'e clonale pour la leuc\'emie lympho\"ide chronique''.

\bigskip

\noindent\textbf{Data availability statement:} No data are associated with this article.

\bibliography{references}

\end{document}